\documentclass{article}

    \PassOptionsToPackage{numbers, compress}{natbib}

\usepackage[preprint]{neurips_2024}

\usepackage[utf8]{inputenc} 
\usepackage[T1]{fontenc}    
\usepackage{hyperref}       
\usepackage{url}            
\usepackage{booktabs}       
\usepackage{amsfonts}       
\usepackage{nicefrac}       
\usepackage{microtype}      
\usepackage{xcolor}         

\usepackage{amsmath,bm}
\usepackage{graphicx}
\usepackage{siunitx}
\usepackage{scalerel}

\usepackage[inline]{enumitem}
\newtheorem{postulate}{Postulate}[section]

\newcommand*{\cX}{\ensuremath{\mathcal{X}}}
\newcommand*{\cD}{\ensuremath{\mathcal{D}}}
\newcommand*{\cR}{\ensuremath{\mathcal{R}}}
\newcommand*{\cC}{\ensuremath{\mathcal{C}}}
\newcommand*{\cL}{\ensuremath{\mathcal{L}}}
\newcommand*{\tx}{\ensuremath{\Tilde{x}}}
\newcommand{\tick}{\scalerel*{\includegraphics{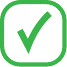}}{X}\ }
\newcommand{\xsymb}{\scalerel*{\includegraphics{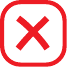}}{X}\ }

\title{Neural rendering enables dynamic tomography}

\author{%
  Ivan Grega\thanks{ig348@cam.ac.uk}, Vikram S. Deshpande \\ 
  University of Cambridge\\
  \And
  William F. Whitney \\
  Google DeepMind \\
}

\begin{document}

\maketitle

\begin{abstract}
  Interrupted X-ray computed tomography (X-CT) has been the common way to observe the deformation of materials during an experiment.
  While this approach is effective for quasi-static experiments, it has never been possible to reconstruct a full 3d tomography during a \textit{dynamic} experiment which cannot be interrupted.
  In this work, we propose that neural rendering tools can be used to drive the paradigm shift to enable 3d reconstruction during dynamic events.
  First, we derive theoretical results to support the selection of projections angles.
  Via a combination of synthetic and experimental data, we demonstrate that neural radiance fields can reconstruct data modalities of interest more efficiently than conventional reconstruction methods.
  Finally, we develop a spatio-temporal model with spline-based deformation field and demonstrate that such model can reconstruct the spatio-temporal deformation of lattice samples in real-world experiments.
\end{abstract}

\begin{figure}[b!]
  \centering
  \includegraphics[width=\linewidth]{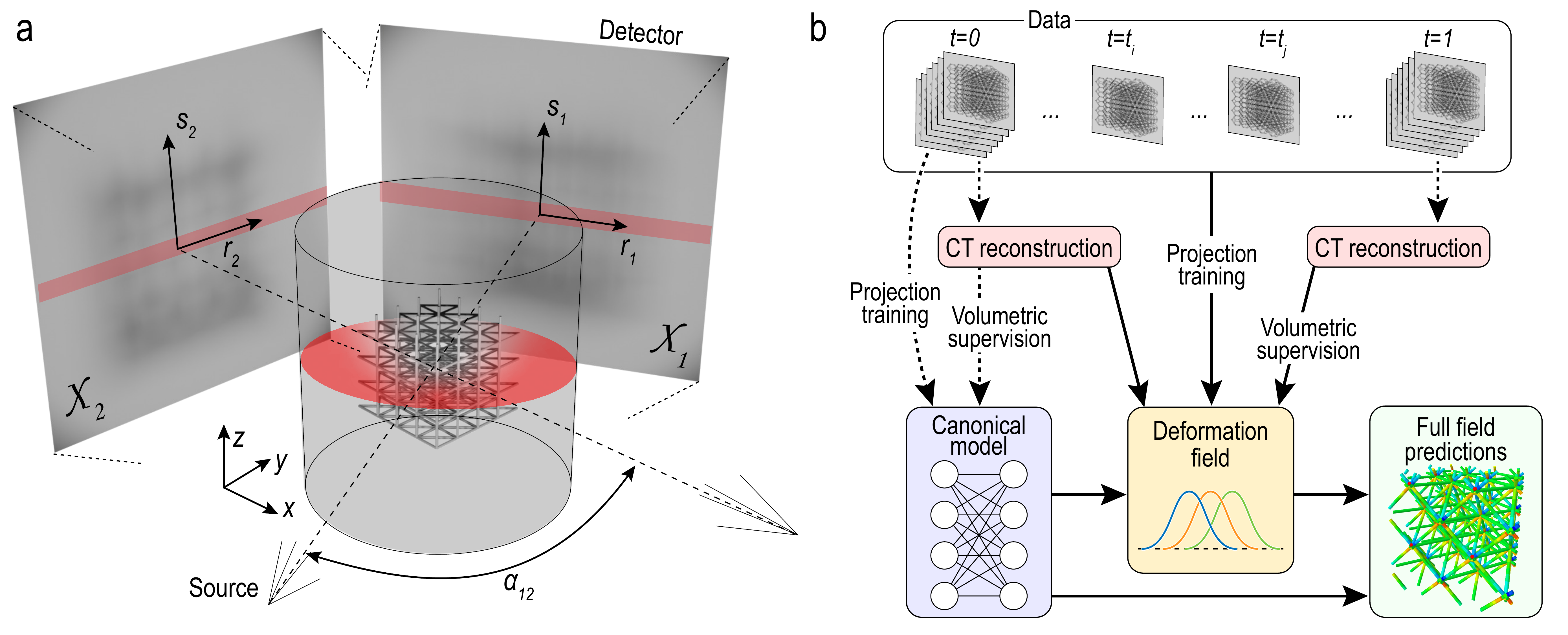}
  \caption{
    {(a)} X-ray CT projection geometry.
    Two projections are illustrated, \(\cX_1,\cX_2\), separated by angle \(\alpha_{12}\). The global reference frame, \((x,y,z)\), and on-screen coordinate frames \((r_i,s_i)\) are indicated.
    {(b)} Illustration of the framework. 
    Projection data of two types are obtained: at times \({t}=0\) and \({t}=1\), rich projection data is obtained, while during the dynamic experiment at intermediate  \({t}\) sparse few-projection data is recorded.
    In pre-training steps, rich many-projection data are reconstructed using computed tomography and the \textit{canonical volume} is fitted (dashed lines).
    Subsequently, the deformation field is calculated from the sparse few-projection data. Full field spatio-temporal information can then be obtained from the two fields.
  }
  \label{fig:proj-geom}
\end{figure}

\section{Introduction}
X-ray computed tomography (X-CT) is an established method used primarily in medical diagnosis and industrial component inspection.
The technique enables a detailed inspection of the portions of  components or tissues which are normally hidden from plain sight.
The operating principle is the acquisition of a large number of X-ray projections from different angles and subsequent reconstruction of these projections into volumetric data.
The tomographic reconstruction is a well-known \textit{inverse problem} and there is a trade-off between the level of detail in the reconstructed volume and the time needed to acquire the projections.
Machine learning techniques have been used to enhance the reconstructed data and reduce noise, thus limiting the radiation dose for patients and achieving a noticeable speedup.
However, there is scope to achieve a complete paradigm shift in the field of experimental science and materials research.
If the number of required projections were reduced to 2 or 3, simultaneous data acquisition from multiple detectors could enable the tomographic reconstruction of dynamic deformation events.

In this work, we demonstrate that it is possible to achieve this paradigm shift by a combination of choosing a suitable prior for the 3d density field and using a differentiable projector/renderer to train this field using only two X-ray projections at intermediate deformation steps.
We first verify our framework with simulated data with a known deformation field.
Then, we demonstrate the use of our framework on a real experimental system of interest - two different lattices undergoing deformation.

As key contributions:
\begin{itemize}[leftmargin=5mm]
    \setlength\itemsep{0em}
    \item We demonstrate that neural radiance fields serve as efficient encoding of tomographic data which enables few-projection reconstruction.
    \item We show that the combination of traditional computed tomography in data-rich regime and neural rendering in data-sparse regime produces a synergistic effect.
    \item We parametrize deformation fields using cubic B-splines with time-modulated weights and demonstrate that it is possible to reconstruct deformation events from as few as two projections.
\end{itemize}


\section{Background}
\subsection{X-ray attenuation tomography}
Imaging based on attenuation of X-rays is an established technique for non-destructive analysis of industrial components and for medical imaging.
The principle of X-ray imaging is Beer-Lambert law for the attenuation of the intensity of electromagnetic radiation passing through a medium.
The differential attenuation of intensity $I$ along path $x$ can be expressed as 
\[
\frac{dI}{dx} = - I \mu(x)
\]
where $\mu$ is the attenuation coefficient of the material. 
This gives rise to the exponential attenuation of signal: 
\(I=I_0 \exp\left(-\int_x\mu(x)dx\right)\).
The attenuation coefficient varies with the energy of X-rays, analogous to how in the visible spectrum different objects absorb different wavelengths of visible light and thus appear as different colors.
However, common detectors only detect the overall aggregated intensity over all wavelengths.

X-ray computed tomography (X-CT) is a technique which enables the reconstruction of 3d volumetric data from 2d X-ray projections. 
We introduce the most important concepts here while more specific details can be found in numerous other texts \cite{2018Industrial,Withers2021X}.
For the simple case of parallel beams, vertical slices through volume can be decoupled and considered separately.
In such case, the density function of projection \(p_\theta(r)\) at angle \(\theta\) can be obtained as an integral through the slice, also known as the \textit{Radon transform}:
\[
p_\theta(r) = \int_{x,y} \mu (x,y) \delta(x \cos\theta+y\sin\theta-r) dx dy
\]
The goal of the tomographic reconstruction is the inverse operation: calculate density \(\mu(x,y)\) from projections \(p_\theta(r)\) which can theoretically be done using the inverse Radon transform. 
However, according to the projection-slice theorem, an infinite number of projections at smallest possible angular increments \(\Delta\theta\rightarrow d\theta\) is needed.
In practice, the number of projections is set such that the maximum displacement of all points of the sample between projections is within one voxel.\footnote{\textit{Voxel} is the 3d equivalent of pixel.}
A typical value for the number of projections is \num{3000}.

\subsection{NeRFs for deformable dynamic scenes}
Numerous frameworks have been developed for the reconstruction of dynamic scenes.
They can be broadly classified on the spectrum between full spatio-temporal networks \((x,y,z,t) \rightarrow (\bm{c},\sigma)\) on the one side and methods which decouple canonical volume \((x_0,y_0,z_0) \rightarrow (\bm{c},\sigma)\) from deformation field \((x,y,z,t) \rightarrow (x_0,y_0,z_0)\) on the other side of the spectrum. \footnote{
    For simplicity we omit viewing directions \((\phi,\theta)\) from the inputs.
}

It has been noted in many works that simply adding time as an input to conventional neural radiance field does not achieve good performance (e.g. \cite{park2021nerfies,pumarola2021d}).
Various types of regularization have been implemented in spatio-temporal networks. 
Examples include depth supervision \cite{xian2021space}
or jointly optimizing a deformation flow field 
\(f(x,y,z,t)\rightarrow (\dot{x}, \dot{y}, \dot{z})\)
\cite{du2021neural,li2021neural}.
While it might be difficult to optimize such models, and many types of loss are needed, they are quite flexible in allowing the topology of the scene to change over time.

A large body of works leverages the concept that in many scenes temporal changes are topology-preserving (\cite{park2021nerfies,guo2022neural,tretschk2021non,liu2022devrf,park2021hypernerf,pumarola2021d}).
These methods train a canonical model which encodes the scene composition at some canonical time \(t_0\) and then find the mapping \((x,y,z,t)\rightarrow(x_0,y_0,z_0)\) for later times \(t\). 
The main distinction between the individual methods is the parametrization of the canonical model and the deformation field.
While most methods use conventional neural radiance fields for the canonical model, some works, e.g. \cite{guo2022neural,liu2022devrf} use voxel grids with trilinear interpolation to query the field at continuous location.
The deformation field \((x,y,z,t)\rightarrow(x_0,y_0,z_0)\) can be parametrized differently in temporal and spatial domain.
Some frameworks (e.g. \cite{park2021nerfies,pumarola2021d,tretschk2021non}) use a general MLP architecture.
Because the optimization problem can be highly ill-posed, various loss terms are imposed to regularize the deformation, for instance elastic energy regularization.
Other methods use a grid fixed in space to represent deformation field.
The parametrization of the temporal evolution of deformation can be discrete using a fixed number of timesteps and linear interpolation \cite{liu2022devrf} or continuous where grid displacements are obtained from an MLP \cite{guo2022neural}.

The drawback of using a canonical model and a deformation field is that while deformation fields can effectively model topology-preserving motion, they fail when model changes topology over time. 
In these cases, it is desirable to have a spatio-temporal network. 
However, optimization with these can be ill-conditioned and there is a need for regularizers. 
Regularization approaches such as elastic regularization can be problematic for topological variations which necessarily require high elastic loss.
Approaches such as low-rank decomposition of the parameters of spatio-temporal neural radiance fields can be used instead \cite{mihajlovic2023resfields,fridovich2023k}. These methods attempt to regularize the problem by finding a lower-dimensional representation of the parameters of spatio-temporal NeRF. 
Work by \citet{park2021hypernerf} is a hybrid approach which uses a deformation model and a canonical model. However, the canonical model is in higher-dimensional hyperspace and a  slice through the hyperspace is selected at each time \(t\).

\section{Related work}

\paragraph{In-situ interrupted tomography}
Deformation in materials is often investigated using interrupted tomography whereby deformation sequence is interrupted at discrete steps to obtain a tomogram.
\citet{shaikeea2022toughness} used this technique for fracture tests of lattices under multiaxial loading while \citet{wang20243d} studied local volume changes in nominally homogeneous rubber specimens under tension, compression and bending. 
Interrupted tomography enables the acquisition of high-quality tomograms, but its main limitation is that the deformation that is applied needs to be quasi-static or the scan can only be done \textit{post-mortem.}

\paragraph{In-operando experiments}
Several efforts have been made to overcome this limitation and study dynamically evolving systems using X-ray imaging.
This is nearly always done on high-performance synchrotron beamlines with specialized detectors which enable high-speed tomography. 
The geometry of the acquisition stage remains the same as for lab-based X-CT: radiation is incident from a single beam through a rotating sample onto a single detector.
Examples include the work by \citet{finegan2015operando} focused on lithium-ion batteries during thermal runaway and a study by \citet{garcia2019using} on metal foaming.
The high-speed sub-second tomographic acquisition requires high rotation rates of the sample which can cause a failure in reconstruction due to the movement inside the sample or, importantly, the high centrifugal forces can change the physical processes that are to be studied.

\paragraph{Few-projection reconstruction}
Several efforts have been made to reconstruct deformation or material model parameters from few radiographs.
Projection-based digital volume correlation (P-DVC) was introduced by \citet{leclerc2015projection} whereby 600 projections are used for the canonical reconstruction of the initial volume and few radiographs are used for the deformed sample of cast iron with graphite spherulites which provide high contrast.
The displacement field is parametrized by nodal displacements of a finite element mesh.
The authors suggest possible strategies for improvement such as elastic regularization, which is used in follow-up work by \citet{taillandier2016measurement}.
The downside of elastic regularization is that a material model has to be assumed \textit{a priori}, which defeats the purpose of DVC if the goal is to find out the material model.
The identification of material model from projection was done by 
\citet{jailin2019fast}. Their sample was a thin quasi-one dimensional dogbone specimen of cast iron with graphite spherulites. The sample was continuously rotated during deformation, completing 1.5 turns over deformation history at an approximate strain rate \SI{2e-4}{s^{-1}} and this sequence of projections was used to calculate the material model.


\paragraph{ML-assisted reconstruction}
A desire to reduce radiation dose in medical imaging has led to the development of numerous ML-assisted reconstruction methods which can operate with a reduced number of projections which 
many scanner manufacturers offer as "deep learning" reconstruction packages.
Typically, such approaches target one of the three domains:
\begin{enumerate*}[label=(\roman*)]
    \item sinogram data,
    \item reconstruction procedure,
    \item tomogram.
\end{enumerate*}
For example, \citet{kim2019extreme} use CNN as sinogram auto-encoder to produce initial guess for an iterative reconstruction algorithm. 
Using such approaches, the required number of projections can reduce to a few hundred and the acquisition time to a few minutes.
While this results in a significant drop in radiation dose, such number of projections/time requirement is still too high for a reconstruction of dynamic experiments.
New avenues can be explored using differentiable rendering tools such as the differentiable projector by \citet{kim2023differentiable}.

\paragraph{NeRF-based approaches}
The use of neural rendering techniques has been demonstrated in X-ray imaging, however most of the approaches rely on pre-training the network on a dataset of objects
\citep{sun2024acnerf,corona2022mednerf,zheng2023ultrasparse}.
This is normally needed because of the ill-posedness of the problem and high number of the degrees of freedom.
However, we wish to circumvent this limitation.
Our goal is to reconstruct dynamic experiments without pre-training the model on pre-existing database and without assuming a material model \textit{a priori}.


\section{Methods}
\subsection{Overview and canonical network}
Our approach uses a familiar scheme which is a combination of a \textit{canonical} network and a \textit{deformation} network (Fig.~\ref{fig:proj-geom}b).
In traditional NeRF methods for 3d scenes, the canonical model takes as input \((x,y,z)\) position and viewing angles, and outputs density \(\rho\), and RGB color. 
A key difference in our work is that X-ray detectors do not distinguish between frequencies of the incident radiation, and the acquired image is monochromatic.
This eliminates the need for color output. 
Similarly, there are no orientation-dependent light reflection effects, which eliminates the need to include viewing direction as input.
Therefore, the canonical network is a density field \({\Theta}(x,y,z) \rightarrow \rho \) with position \((x,y,z)\)  as input and density \(\rho\) as output.
We use multi-resolution hash encoding \cite{muller2022instant} as \({\Theta}\).
A key difference between our framework and conventional X-CT reconstruction is that in classical X-CT methods the density field is encoded as 3d voxel grid, whereas in this work it is represented as a continuous function \({\Theta}(x,y,z)\).

\subsection{Spline-based deformation network}
The deformation field is based on cubic B-splines by \citet{rueckert1999nonrigid}.
We modified the method to account for temporal variations in deformation.
Let us denote time-varying displacement field \(\bm{u}(\bm{x},t)\) with its three components, \(u_x,u_y,u_z\).
We parametrize the field using cubic \textit{B-splines}.
B-spline field is a scalar field defined by \(n_x\times n_y\times n_z\) grid points and a weight \(w\) at each grid point. 
The 3 components of displacement field,
\(u_x,u_y,u_z\), are considered as 3 independent scalar fields.
The value of displacement component \(u\), e.g. \(u_x\), at point \((x,y,z)\) is given by interpolation using weights of the four closest grid points along each dimension:
\[
u(x,y,z,t)=\sum_{l=0}^3\sum_{m=0}^3
\sum_{n=0}^3 B_{l}(\Tilde{x})B_{m}(\Tilde{y})B_{n}(\Tilde{z}) w(t)_{i_x+l, i_y+m, i_z+n}
\]
where \(B_i(x)\) is fixed cubic B-spline function,
\(i_x,i_y,i_z\) are indices of the relevant grid points, \(\Tilde{x},\Tilde{y},\Tilde{z}\) are modulo-subtracted coordinates,
and \(w(t)_{i_x, i_y, i_z}\) is the trainable weight of the corresponding spline control point. 
Full details of the interpolation scheme are explained in the Appendix.

We encode temporal variation of the field by time-varying weights \(\bm{w}(t)\) on a fixed grid (i.e. grid control points remain fixed in space). 
We derive weights from a MLP \(\phi(t) \rightarrow \bm{w}(t)\) which takes time \(t\) as input and returns the complete set of weights \(\bm{w}(t)\).
See Appendix for more details.
The key points to note are:
\begin{enumerate*}[label=(\roman*)]
    \item the field can be queried at arbitrary continuous position and time;
    \item the displacement at any query point is a linear function of the weights of the 64 closest grid points.
\end{enumerate*}

\subsection{Attenuation-based rendering}
The most fundamental difference between X-ray imaging and photographic imaging from the perspective of neural rendering is the rendering equation.
The majority of neural rendering frameworks uses a discretized version of the equation \(C = \int T(t)\sigma(t) c(t) \) with 
transmittance \(T(t) = \exp\left( -\int \sigma(s) ds \right)\)
integrated from the observer to coordinate \(t\).
Such model works very well in the classical photographic image setting where the observed color is only emitted from a thin surface layer of objects.
However, in X-ray attenuation tomography, rays of radiation emitted from the source penetrate the object to reach the detector. 
The observed intensity at the detector, \(I\), is the source intensity, \(I_0\), attenuated by the object: \(I = I_0 \exp \left(-\int \mu(s) ds \right)\). 
Note that there is symmetry along the ray -- unlike in the photographic setting -- and reversing near and far bounds of the ray does not change the detected intensity.
We implement attenuation renderer that calculates pixel intensities based on this equation.
The three RGB channels are kept the same as the attenuation projection is monochromatic.

The following additional steps are taken to adapt the NeRF method to X-ray data.
\begin{enumerate*}[label=(\arabic*)]
    \item In most lab-based X-CT systems, the detector is stationary and the sample rotates around an axis of rotation.
    In this work, we conceptually fix the sample at the origin and rotate the camera around the origin (Fig.~\ref{fig:proj-geom}a).
    \item During X-ray tomography, the parameters of the beam (energy and current) and scan (exposure time) are configured depending on the examined sample.
    Notably, the \textit{flat field} image \footnote{A flat field image is an image acquired by the detector in absence of any objects in the field of view.}
    is not fully saturated (white) but appears grey.
    To capture this behaviour, we add a trainable flat field background to the model.
\end{enumerate*}
Details are explained in Appendix~\ref{sec:app-flat-field}.


\begin{figure}
  \centering
  \includegraphics[width=\linewidth]{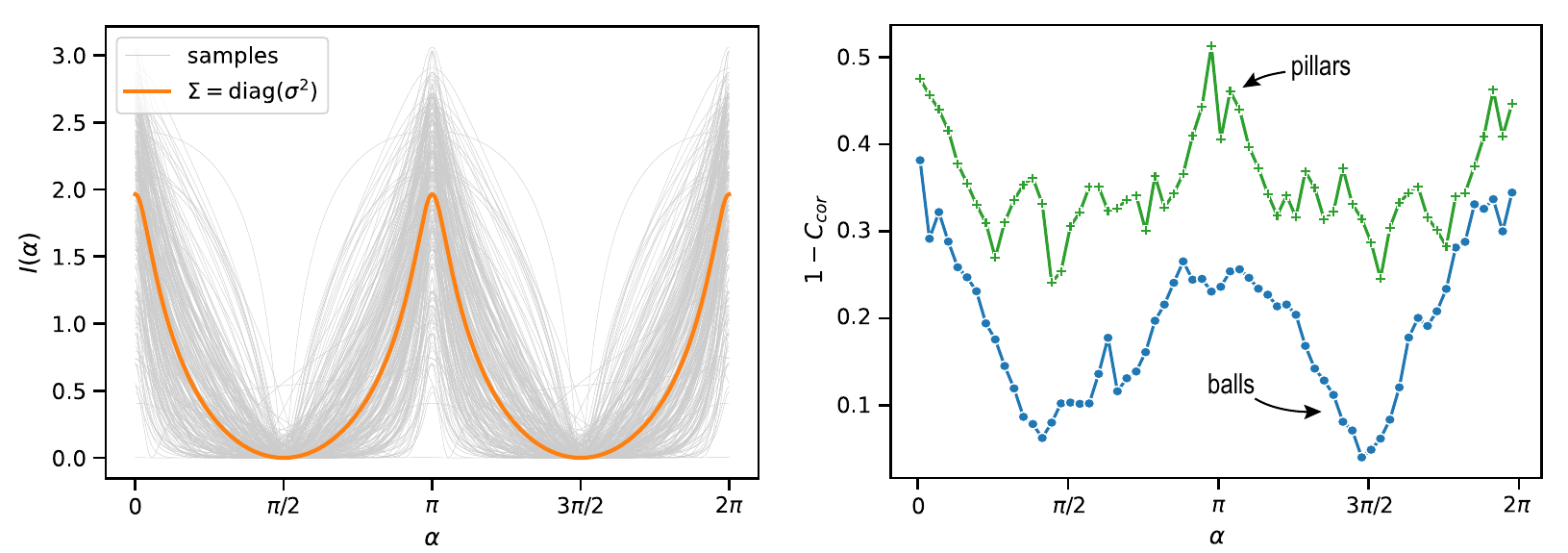}
  \caption{
  {(a)} 
  Theoretical result for the mutual information between two projections separated by angle \(\alpha\). 
  Thin grey lines correspond to samples drawn from various Gaussian distributions while the thick orange line is the mutual information for an axisymmetric Gaussian.
  {(b)}
  Reconstruction error plotted as \(1-\mathcal{C}_{cor}\) as a function of separation angle  \(\alpha\) between 2 projections which are used to train the model for two simulated datasets (\textit{balls} and \textit{pillars}). }
  \label{fig:val-inf-theor}
\end{figure}

\subsection{Information-theoretic data collection}\label{sec:mutual-inf-methods}
Here we approach the question "How do we choose the optimum projection angles?" from the perspective of information theory.
Let us define random variable \(\cD\) to represents the underlying data in 3d volume.
We observe \(N\) 2d projections of this 3d data \(\cX=\cX_1,...,\cX_N\) at angles \(\theta_1,...,\theta_N\). 
In Figure~\ref{fig:proj-geom}a we illustrate the setup with 2 projections, \(\cX_1,\cX_2\).

\begin{postulate}
The optimum set of angles $\theta_i$ is such that the entropy of data given all projections \(H(\cD|\cX)\) is minimized:
\[
H(\cD|\cX) = H(\cD) + H(\cX|\cD) - H(\cX)
\]
\end{postulate}
The term \(H(\cD)\) represents the complexity of the 3d volume.
The second term \(H(\cX|\cD)\) depends only on noise level in projection acquisition.
The final term, \(H(\cX)\) is the one that depends on the choice of projection angles.
In particular, we make the assumption that higher-order mutual information can be approximated as a sum of pairwise terms. Then
\[
H(\cX) = \sum_i\,H(\cX_i) + \sum_{i,j}I(\cX_i,\cX_j)
\]

With the stationarity assumption, the mutual information for two projections should only depend on their angular separation \(\alpha_{ij}=|\theta_i-\theta_j|\). 
We use Gaussian priors on \(\cD\) to obtain an analytical expression for pairwise mutual information :
\begin{equation}
    I(\alpha_{ij}|\bm{\Sigma})= 
        -\frac{1}{2} \ln{\left( 1- \frac{
    \left(\bm{b_i}^T\bm{\Sigma}\bm{b_j} \right)^2}{
    (\sigma_0^2+\bm{b_i}^T\bm{\Sigma}\bm{b_i})(\sigma_0^2+\bm{b_j}^T\bm{\Sigma}\bm{b_j})
    }
    \right)
    }
\end{equation}
where \(\sigma_0\) is the noise level in \(p(\cX|\cD)\), \(\bm{\Sigma}\) is the covariance matrix of the Gaussian and direction vectors \(\bm{b_i}=[ 1, 0 ]^T\), \(\bm{b_j}=[ \cos\alpha_{ij}, \sin\alpha_{ij} ]^T\).
In the case of an axisymmetric Gaussian, the formula reduces to:
\begin{equation} \label{eq:mut-inf-simple}
    I(\alpha_{ij}) = -\frac{1}{2}\ln{\left(1 - \frac{\cos^2\alpha_{ij}}{(1+\varepsilon)^2}\right)}
\end{equation}
where \(\varepsilon\) is a nondimensional positive number that characterizes the level of noise in the projection-generating forward model.
We provide a full derivation in the Appendix. 
Examples of \(I(\alpha_{ij}|\bm{\Sigma})\) for various choices of \(\bm{\Sigma}\) as well as the simplified expression for axisymmetric Gaussian are plotted in Figure~\ref{fig:val-inf-theor}a.
We observe that \(I\) has minima at \(\alpha=\{\pi/2, 3\pi/2\}\) which implies that the optimum X-ray angles for 2 images are \(\{0, 90\}^\circ\).
Note the difference from photographic NeRFs -- 
in the photographic setting, the camera can only see one side of an object, so with 2 images it is natural to separate them by \ang{180}.
The simplified form in (Eq.~\eqref{eq:mut-inf-simple}) enables us to numerically optimize projection angles for any given number of projections.
Throughout this work, we use such optimum projection angles unless otherwise stated.

\subsection{Definitions of metrics} \label{sec:metrics}
In neural rendering setting for 3d scenes, it is natural to calculate error based on withheld views of the scene. A common metric is Peak Signal-to-Noise Ratio \textit{(PSNR)}:
\begin{equation}
\text{PSNR}(y, \hat{y}) = 10  \log_{10} \left(\frac{\max(y)^2}{\left< (y-\hat{y})^2 \right>}\right)
\end{equation}
where \(y\) is the pixel value of the target image, \(\hat{y}\) is the noisy approximation, and angled brackets \(\left< \cdot \right>\) denote average taken over the image.
We use this metric in cases when we do not have access to high-quality reconstructions of volumes.

When rich projection data is available, we use conventional X-CT reconstruction to obtain 3d voxel data for the density field.
However, the scale of density values obtained from X-CT reconstruction is arbitrary.
Therefore, we need a metric that is sensitive to local and global shape variations, but insensitive to global scaling or shift in density values.
Normalized correlation coefficient \(\cC_{cor}\) is such metric:
\begin{equation}
\mathcal{C}_{cor}(y, \hat{y}) = \frac{
    \sum_{i\in \Omega} \left(y_i-\left<y\right>\right) \left(\hat{y}_i-\left<\hat{y}\right>\right)
    }{
    \sqrt{\sum_{i\in \Omega} \left(y_i-\left<y\right>\right)^2 \sum_{i\in \Omega} \left(\hat{y}_i-\left<\hat{y}\right>\right)^2}
    }
\end{equation}
where \(y\) and \(\hat{y}\) are the target and fitted density values, respectively, sums are taken over voxels \(i\) in domain \(\Omega\), and quantities in angled brackets \(\left< \cdot \right>\) are averaged values over this domain.


For experiments with simulated deformation, suppose displacement vector is \(\bm{u}(\bm{r},t)\), where \(\bm{r}\) is the position vector \([x,y,z]^T\), and the corresponding ground truth is \(\bm{\hat{u}}(\bm{r},t)\).
Displacement error \(\mathcal{E}_{disp}\) is defined as the spatial and temporal average of the squared L2 norm of the difference of the predicted and true displacement vectors:
\[
\mathcal{E}_{disp} = \int_0^1 \mathrm{d}t \frac{1}{|\Omega|} \int_\Omega \| \bm{u}(\bm{r},t) - \bm{\hat{u}}(\bm{r},t) \|^2 \mathrm{d}\Omega
\]
In practice, we approximate the integrals by discrete sums.

\subsection{Volumetric supervision} \label{sec:volumetric-supervision}
Let us define dimensionless time \({t}\) for an experiment such that \({t}=0\) denotes the start of the experiment and \({t}=1\) denotes the end.
While we often have to work in the sparse few-projection regime at intermediate times \(0<{t}<1\), it is straightforward to obtain rich many-projection data at the start and end of the experiment \({t}\in\{0,1\}\).
Therefore, we can reconstruct the volumetric data in the rich regime using conventional tomography and use this information to improve the training of the canonical model and deformation field.
Since the scales for density levels may differ in the canonical volume and the CT-reconstructed volume, we use normalized correlation coefficient \(\cC_{cor}\) as target accuracy metric.
We define the associated \textit{volumetric loss}, \(\cL_{vol}\), as \(\cL_{vol}=- \cC_{cor}\).

\begin{figure}
  \centering
  \includegraphics[width=\linewidth]{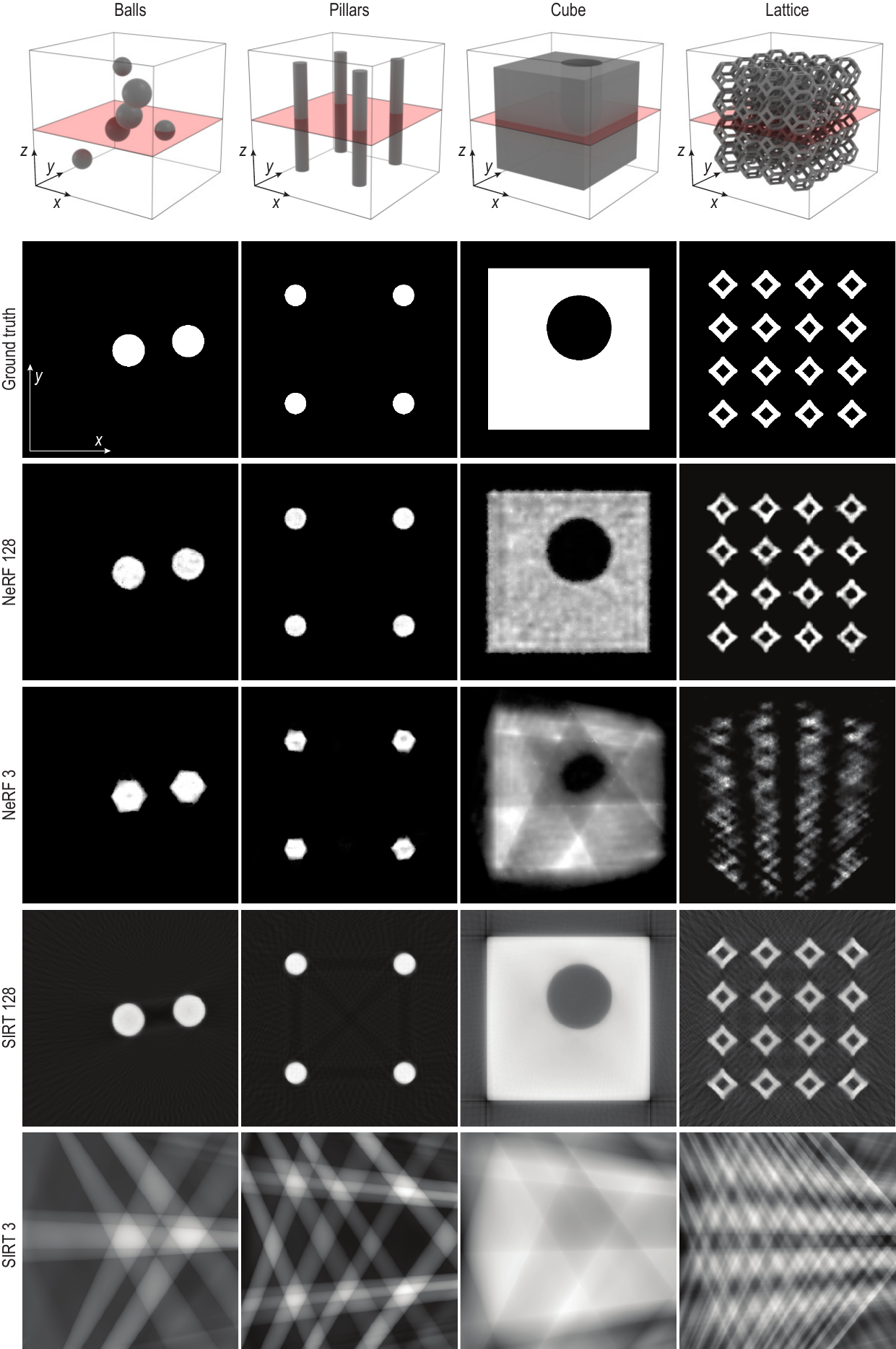}
  \caption{Overview of the data efficiency study. Each column corresponds to one dataset. An \(xy-\)plane is highlighted in the 3d view which corresponds to the slicing plane shown in the subsequent rows: ground truth, NeRF trained with 128 and 3 projections, SIRT (conventional reconstruction) trained with 128 and 3 projections.}
  \label{fig:data-overview}
\end{figure}

\subsection{Data}
All data used in this work was experimentally captured or computationally generated by us.
Four simulated datasets were generated for Section~\ref{sec:static-reconstructions}: \textit{balls, pillars, cube, lattice} (see Fig.~\ref{fig:data-overview}). 
For each object, the training set comprised of \num{256} equispaced projections in the horizontal plane, while
the test set was composed of \num{25} further projections out of plane.
Dataset for simulated deformation experiment (Section~\ref{sec:simulated-def-exp}) was generated using the \textit{lattice} object and an additional simulated deformation field. 
Further details including a link to the scripts used to generate these projections can be found in the Appendix.
Experimental datasets for use in Section~\ref{sec:static-reconstructions} were collected in our lab-based X-CT system to mimic the simulated datasets as closely as possible.
For real-world deformation experiments (Section~\ref{sec:real-def-exp}), nominally 1024 projections were collected for high-fidelity X-CT reconstruction.
A subset of 128 equispaced projections was saved for NeRF reconstruction.
At intermediate times, 12 equispaced projections were saved, but only 2 were used in this work.

\section{Experiments}
The experiments in the paper are split into two parts:
\begin{enumerate}[label=(\arabic*)]
    \item Static reconstructions: 
    \begin{enumerate*}[label=(\roman*)]
        \item we establish using 4 experimental and simulated datasets that NeRFs are an efficient reconstruction tool;
        \item we verify the theoretic results about optimum projection angles based on mutual information;
        \item we show that combining projection data with volumetric X-CT data has a synergistic effect.
    \end{enumerate*}
    \item Deformation reconstructions: 
    \begin{enumerate*}[label=(\roman*)]
        \item we verify our method with simulated deformation where the displacement field is known;
        \item we deploy the method to two real-world deformation experiments with complexities in deformation field and topology.
    \end{enumerate*}
\end{enumerate}

\subsection{Static (canonical) reconstructions} \label{sec:static-reconstructions}

\subsubsection{Data efficiency of neural rendering} \label{sec:data-efficiency}
To establish how well the neural rendering method works for the X-ray data modality, we first train neural radiance fields on simulated projections through static objects.
To characterize accuracy, we use normalized correlation coefficient \(\mathcal{C}_{cor}(y,\hat{y})\) as defined in section~\ref{sec:metrics}.
We numerically evaluate the normalized correlation coefficient over a grid of \(200\times 200 \times 200\) samples.
The target data \(y\) comes from our explicit knowledge of the object density in simulated datasets.

\begin{figure}
  \centering
  \includegraphics[width=\linewidth]{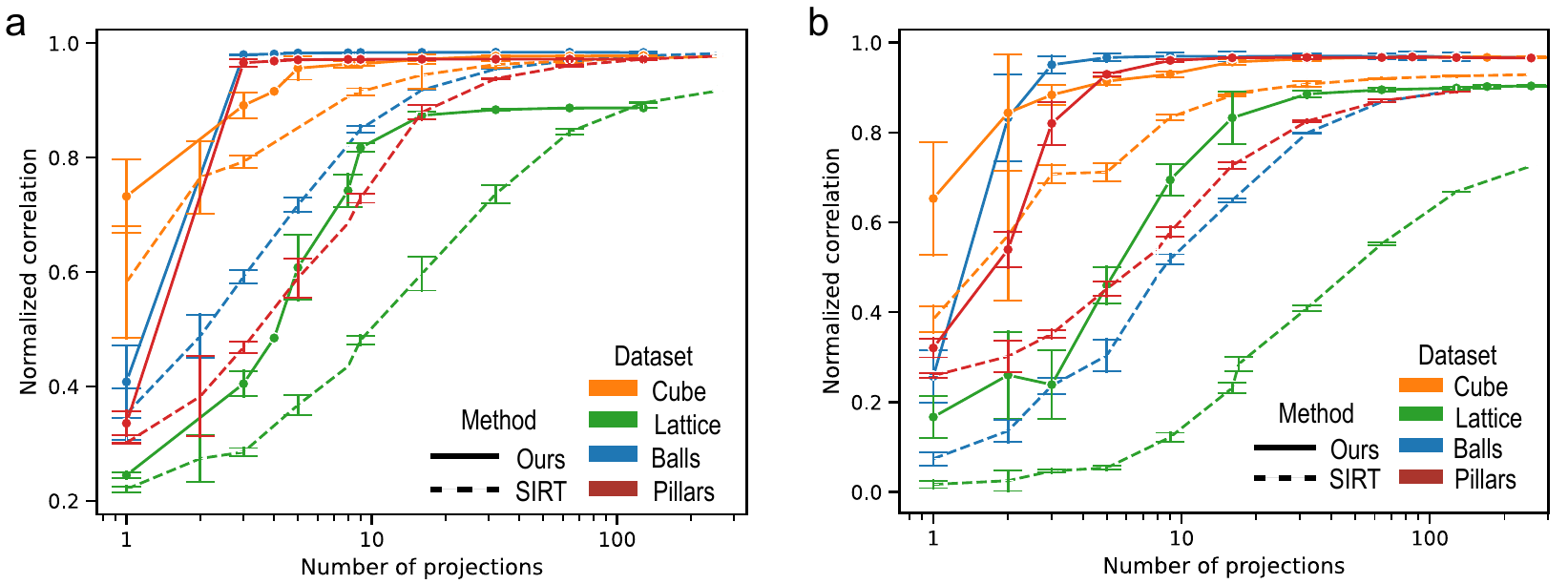}
  \caption{(a) Convergence of normalized correlation \(\cC_{cor}\) with number of projections for various datasets. 
  Our neural rendering technique is more efficient up to approx. 100 projections.
  (b) Convergence of normalized correlation \(\cC_{cor}\) with number of projections for various datasets using experimental data.}
  \label{fig:loss-scaling}
\end{figure}

In Figure~\ref{fig:loss-scaling}a and Table~\ref{tab:norm-correlation}, we present the normalized correlation coefficient \(\cC_{cor}\) as a function of the number of observed projections for various datasets and reconstruction methods.
Our neural rendering method is compared against two conventional reconstruction methods: \textit{SIRT}, and \textit{CGLS}.
For clarity, the figure only shows SIRT and NeRF, as  SIRT outperforms CGLS across the spectrum.
Table~\ref{tab:norm-correlation} includes a comparison of all three methods.

We also collected experimental data with analogous objects and evaluated the convergence of normalized correlation with the number of projections (Figure~\ref{fig:loss-scaling}b).
The behaviour for experimental data is very similar to simulated data, which supports the case that our simulated data generation process is an appropriate model of real X-ray projections.

In summary, we make the following observations.
\begin{enumerate}[label=(\arabic*),leftmargin=*,wide=0pt]
    \item In the data-sparse regime, our neural rendering method outperforms conventional reconstruction algorithms for all examined datasets.
    To achieve comparable normalized correlation when using conventional algorithms, the number of required projections is larger by approximately a factor of 3.
    
    \item The quality of reconstruction is a strong function of the complexity of the underlying data.
    The \textit{balls} and \textit{pillars} datasets are the easiest to reconstruct and with approximately 3 projections the neural rendering method captures the data with correlation coefficient \(\mathcal{C}_{cor}=0.98\).
    For the \textit{cube} dataset approximately 6 projections are needed for satisfactory reconstruction. 
    The most challenging dataset to reconstruct is the \textit{lattice} dataset. Our neural rendering method does not produce a correlation larger than \(\mathcal{C}_{cor}=0.9\).

    \item The conventional reconstruction algorithms match the neural rendering method at approximately 100 projections, and in the data-rich regime they are able reconstruct even very complex data (e.g. lattice dataset) extremely well.
    However, the reconstruction of experimental lattice data is more challenging and full conventional tomographic acquisition (with approximately 1000 projections) is needed.
\end{enumerate}

\begin{table}[]
    \centering
    \resizebox{\textwidth}{!}{%
    \begin{tabular}{ccccccccccccc}
    \toprule
    {} & \multicolumn{3}{c}{Balls} & \multicolumn{3}{c}{Cube} & \multicolumn{3}{c}{Lattice} & \multicolumn{3}{c}{Pillars} \\
    {} & 3 & 9 & 256 & 3 & 9 & 256 & 3 & 9 & 256 & 3 & 9 & 256 \\
    \midrule
    Ours & \textbf{0.98} & \textbf{0.98} & \textbf{0.98} & \textbf{0.89} & \textbf{0.96} & \textbf{0.98} & \textbf{0.40} & \textbf{0.82} & 0.91 & \textbf{0.97} & \textbf{0.97} & \textbf{0.98} \\
    CGLS & 0.57 & 0.83 & 0.97 & 0.74 & 0.79 & 0.97 & 0.01 & 0.44 & 0.87 & -0.01 & 0.73 & 0.96 \\
    SIRT & 0.59 & 0.85 & \textbf{0.98} & 0.79 & 0.92 & \textbf{0.98} & 0.29 & 0.48 & 0.92 & 0.47 & 0.73 & \textbf{0.98} \\
    \bottomrule\vspace{1mm}
    \end{tabular}
    }
    \caption{
    Volumetric accuracy calculated as normalized correlation for each dataset as a function of various numbers of training projections.
    We compare our neural rendering technique to conventional reconstruction algorithms.
    }
    \label{tab:norm-correlation}
\end{table}

\subsubsection{Validation of information-theoretic result}
We run a series of experiments to validate the proposed surrogate metric for pairwise mutual information \(I(\alpha_{ij})\) from section~\ref{sec:mutual-inf-methods}. 
First we run training with input data of 2 projections and observe the normalized correlation metric \(\mathcal{C}_{cor}\) as a function of the angular separation between the two projections, \(\alpha_{01}=\alpha\).
We choose two datasets, balls and pillars, as these datasets achieve near-perfect reconstruction from 2 optimally-chosen projections.
In Figure~\ref{fig:val-inf-theor}b we plot \(1-\cC_{cor}\), which should be zero for perfect reconstruction. 
It is evident that the minimum is obtained around \(\alpha_{ij}\in\{\pi/2, 3\pi/2\}\) and the shape of the curve resembles the graphs of analytical closed-form expressions for pairwise mutual information in Fig.~\ref{fig:val-inf-theor}a. 

\begin{figure}
  \centering
  \includegraphics[width=\linewidth]{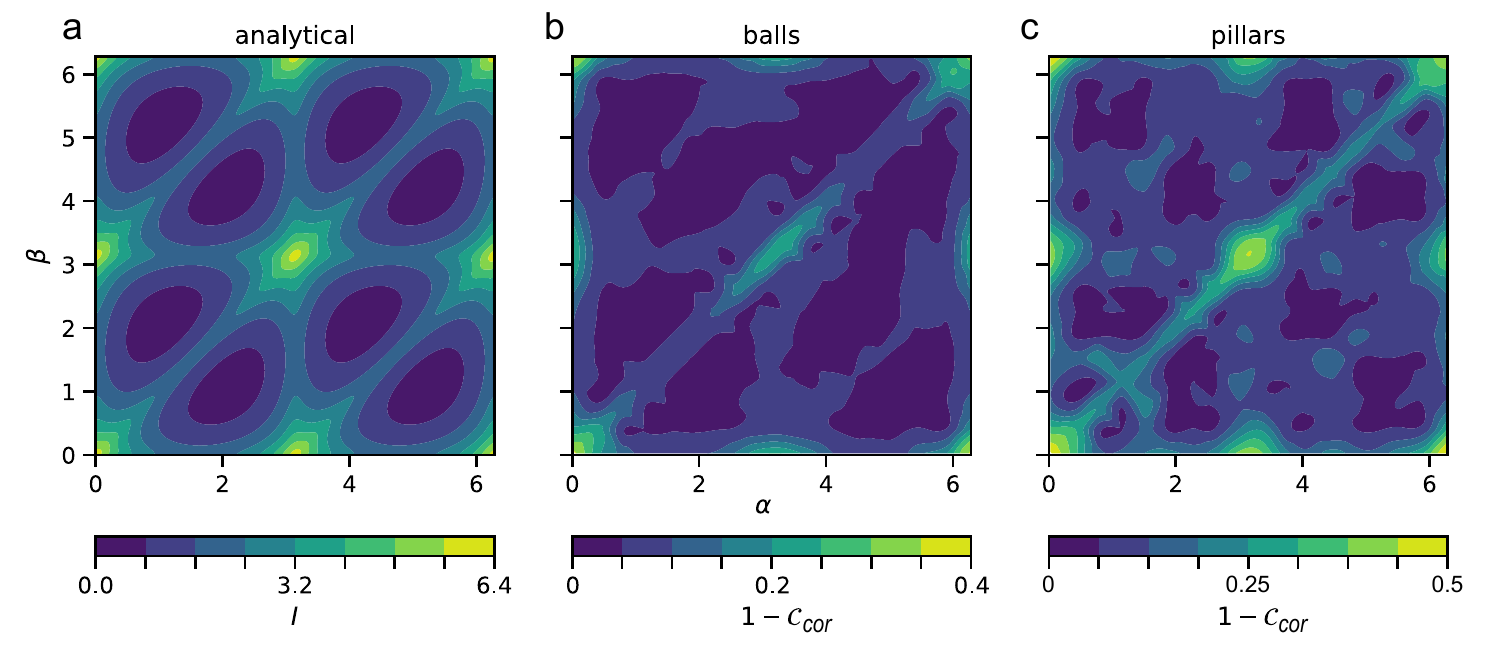}
  \caption{
  Verification of mutual information theoretical result with three projections.
  {(a)} Analytical contours of the mutual information for three projections.
  (b,c) Contours of reconstruction error proxy, \(1-\cC_{cor}\), for three projections in simulated datasets: balls (b) and pillars (c).
  }
  \label{fig:val-inf-theor-3}
\end{figure}

Further verification is done for the case of three projections.
In Figure~\ref{fig:val-inf-theor-3}a we plot the analytical expression for mutual information \(I=I(\cX_0;\cX_1)+I(\cX_0;\cX_2)+I(\cX_1;\cX_2)\) under the assumptions of pairwise interactions and axisymmetric Gaussian data distribution.
We define angles \(\theta_0=0\), such that \(\alpha_{01}=\theta_1 - \theta_0=\alpha\) and \(\alpha_{02}=\theta_2 - \theta_0 = \beta\).
The optima are \(\alpha\simeq \pi/3, \, \beta\simeq 2\pi/3\).
In Fig.~\ref{fig:val-inf-theor-3}b,c we plot a metric for reconstruction error \(1-\cC_{cor}\).
We observe that the contours of \(1-\cC_{cor}\) resemble the contours of \(I\), which supports the postulate that optimum projection angles correspond to the angles which minimize the mutual information between the projections.

\subsubsection{Volumetric supervision to guide NeRF field}
\begin{figure}
  \centering
  \includegraphics[width=\linewidth]{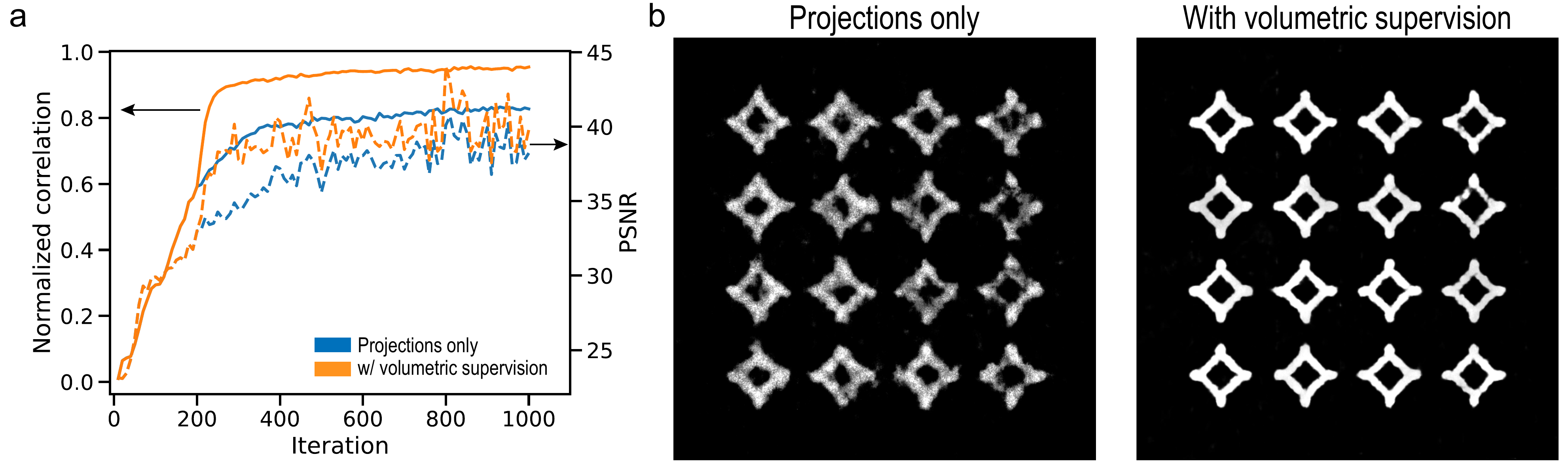}
  \caption{Introduction of volumetric supervision with high-fidelity X-CT data improves reconstruction.
  {(a)} Normalized correlation (solid lines) and peak signal to noise ratio (PSNR, dashed lines) with and without volumetric supervision.
  {(b)} Slices from the canonical volume at the midplane \(z=0\) trained purely using projections, and with volumetric supervision.
  }
  \label{fig:volumetric-supervision}
\end{figure}

We have shown that for complex objects, the neural rendering method does not achieve normalized correlation coefficient greater than \(\mathcal{C}_{cor}=0.9\) while conventional data-rich X-CT can reconstruct such volumes with high fidelity when many projections are available.
We leverage this observation to improve the quality of the canonical volume.
We propose a hybrid method of reconstruction when high-fidelity X-CT can be obtained (\(t=0\) and \(t=1\)).
Volumetric loss based on normalized correlation \(\mathcal{C}_{cor}\) between the predicted density of the canonical volume and the ground truth X-CT volume is calculated and added to the training loss: \(\mathcal{L} = \mathcal{L}_{proj} + \lambda \mathcal{L}_{vol}\). 
In Figure~\ref{fig:volumetric-supervision}a, we show the comparison of normalized correlation and peak signal to noise ratio for the simulated lattice dataset trained with and without volumetric supervision.
Note that for the case with volumetric supervision, the coefficient \(\lambda\) was zero for the first \num{200} iterations and then changed to \(\lambda=0.005\). 
In Figure~\ref{fig:volumetric-supervision}b we show corresponding cross-sectional slices.
It is clear that the introduction of volumetric supervision improves both the metrics and also perceptual quality of reconstruction.

\subsection{Deformation reconstructions}
\subsubsection{Reconstruction of simulated deformation experiment} \label{sec:simulated-def-exp}
In an expriment with simulated deformation field, we construct simulated projections for a \(4\times4\times4\) Kelvin lattice.
We use \num{256} projections at \(t=0\) and \(t=1\), and 2 projections for nine intermediate deformation steps \({t}=0.1,...,0.9\).
The applied deformation field is inspired by uniaxial tensile test, but we add an additional \(xy\) term to increase the complexity:
\begin{align*}
    &\qquad u_x=0 \qquad u_y=0 \qquad u_z= \frac{0.05t}{
    1+\exp{\left( - \frac{z-0.05{t}-xy} {0.15} \right)}
    }
\end{align*}

\begin{figure}
  \centering
  \includegraphics[width=\linewidth]{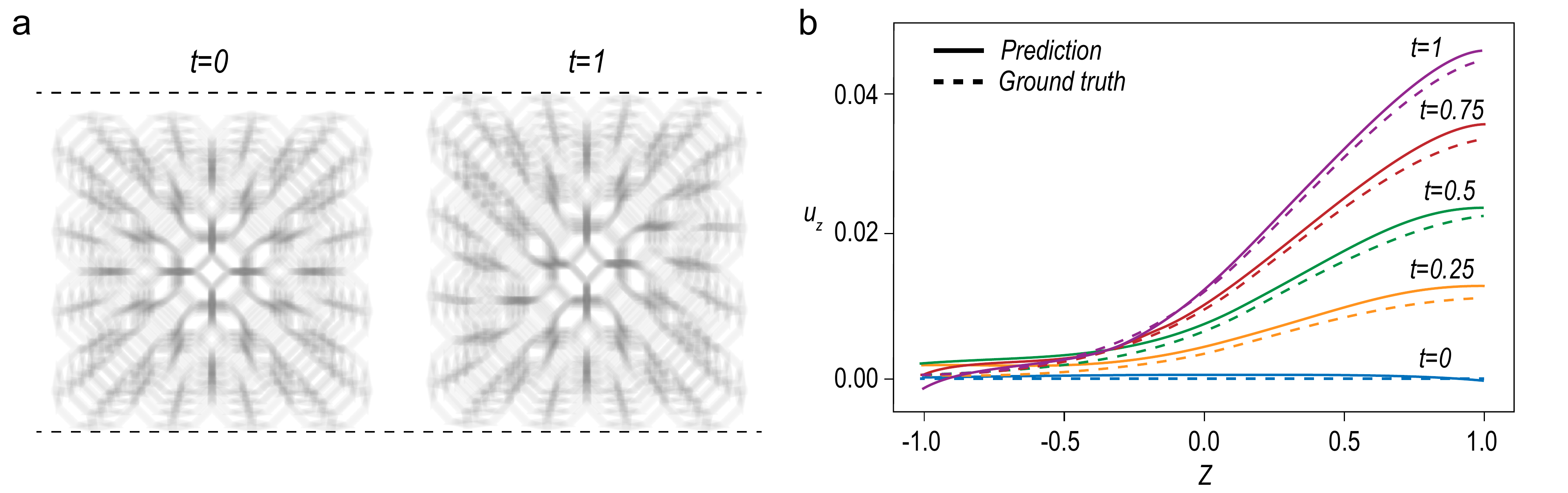}
  \caption{ Experiments with simulated deformation. 
  {(a)} Projections at \(t=0\) and \(t=1\). Dashed lines indicate the vertical positions of the top and bottom of the sample at the onset of deformation.
  {(b)}  Displacement \(u_z\)
  on a vertical line through the specimen \((x=y=-0.5)\) as a function of \(z\) for various times \({t}\).
  Ground truth functions shown as dashed lines.
  } 
  \label{fig:simulated-reconstruction}
\end{figure}

We show projections for \(t=0\) and \(t=1\) in Figure~\ref{fig:simulated-reconstruction}a.
The canonical volume is reconstructed at \(t=0\) using the hybrid method: loss is a projection loss based on 256 projections combined with volumetric loss based on high-fidelity X-CT.
Subsequently, deformation field is trained with B-spline grid resolution of \(6 \times 6 \times 6\).
A comparison of the predicted and ground truth deformation field is shown in Figure~\ref{fig:simulated-reconstruction}a where
we plot component  \(u_z\) along a vertical line through the specimen \((x=y=-0.5)\).
The predicted displacement field matches the imposed field with a high degree of accuracy.

We carry out an ablation study to consider the effect of removing two components of the training pipeline: 
\begin{enumerate*}[label=(\arabic*)]
    \item volumetric supervision, and
    \item rich data at \(t=1\).
\end{enumerate*}
In Table~\ref{tab:simulated-reconstruction} we tabulate displacement error, \(\mathcal{E}_{disp}\) (as defined in Section~\ref{sec:metrics}) for four cases.
It can be observed that the two components improve the accuracy of reconstructed displacement field both individually and in synergy.

\begin{table}[]
    \centering
    \begin{tabular}{ccccc}
    \toprule
    Volumetric supervision & \xsymb & \tick & \xsymb & \tick \\
    Rich data \(t=1\) & \xsymb & \xsymb & \tick & \tick \\
    \(\mathcal{E}_{disp}\) & \num{6.4}& \num{5.1} & \num{2.1}  & \(\num{0.5}\) \\
    \bottomrule \\
    \end{tabular}
    \caption{
    Displacement error \(\mathcal{E}_{disp} [\times 10^{-5}]\)
    }
    \label{tab:simulated-reconstruction}
\end{table}

\subsubsection{Reconstruction of real-world deformation experiment} \label{sec:real-def-exp}
\begin{figure}
  \centering
  \includegraphics[width=\linewidth]{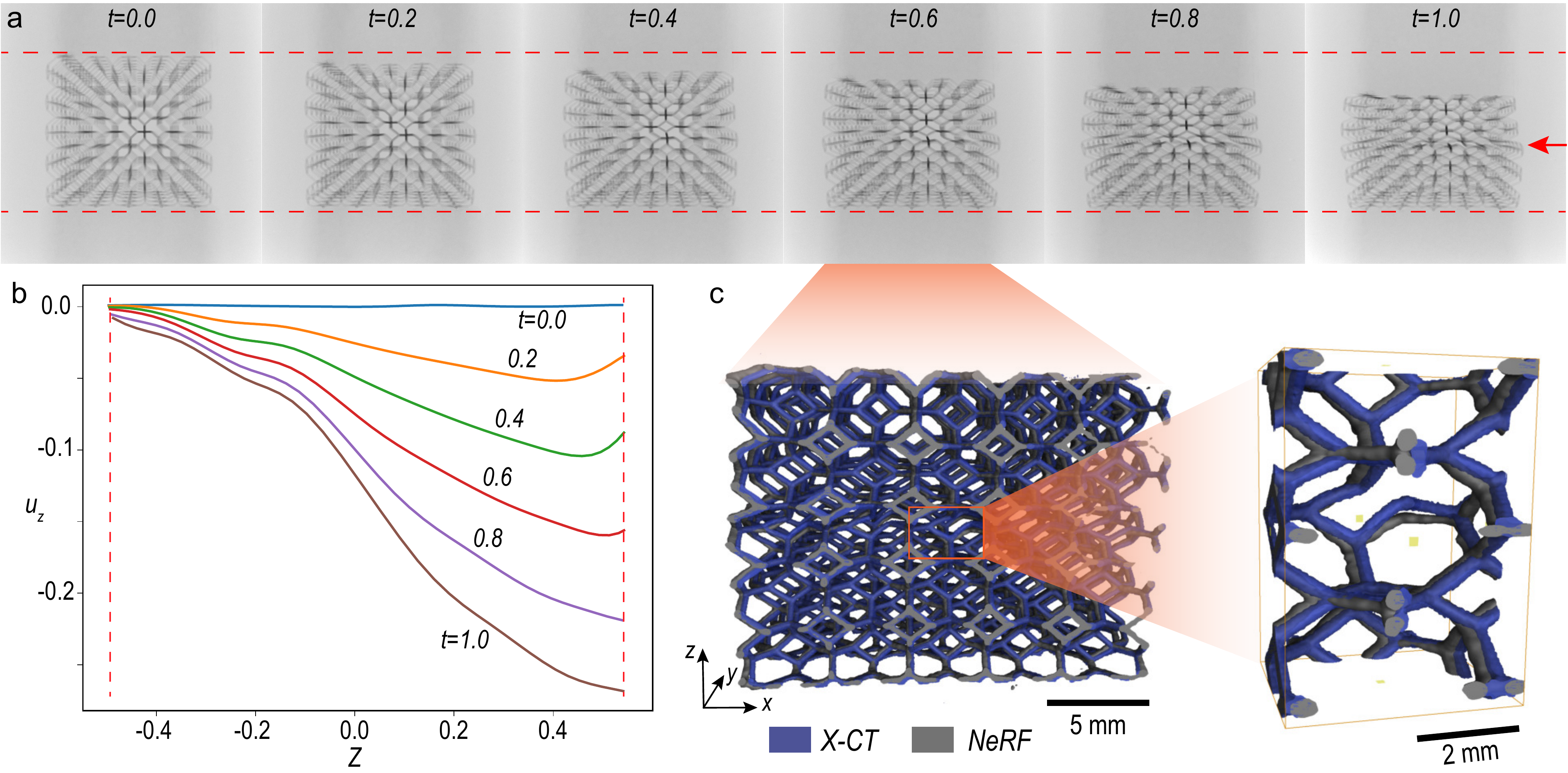}
  \caption{ 
  Reconstruction of localizing Kelvin deformation experiment.
  {(a)} Radiographs at 6 selected timestamps. Arrow indicates deformation localization.
  {(b)} Calculated displacement field \(u_z(x=0,y=0,z)\).
  {(c)} 3d renderings of the reconstructed volume at \(t=0.6\) compared with withheld tomogram.
  } 
  \label{fig:kelvin-recon}
\end{figure}

\paragraph{Kelvin foam with planar localization}
In Figure~\ref{fig:kelvin-recon} we show the reconstruction of a real \(5\times5\times5\) Kelvin lattice.
The sample was loaded in compression with a linear mapping between remote displacement \(\Delta\) and reconstruction time \(t\): \(\Delta / \si{mm} = 5t \).
Six projections at various timesteps \(t\) are shown in Fig.~\ref{fig:kelvin-recon}a with red dashed lines indicating the vertical positions of the top and bottom of the sample at the onset of deformation.
The red arrow indicates the vertical position of the localization of the deformation.

The inferred deformation field is presented in Figure~\ref{fig:kelvin-recon}b, where we plot vertical displacement, \(u_z\), along the centerline of the sample \((x=y=0)\).
While in the previous section it was possible to compare the inferred displacement field with the ground truth, in this case we do not have access to the true displacement field.
However, we recorded full tomograms at several intermediate timesteps.
Therefore, we can visualize the difference between the high-fidelity reconstructions and the our NeRF reconstructions. 
In Figure~\ref{fig:kelvin-recon}c, we show the reconstruction of the sample at \(t=0.6\) with clipping applied at the midplane \((y=0)\).
Both the high-fidelity X-CT reconstruction (blue) and the NeRF reconstruction (grey) are superimposed. 
An inset shows a small central portion of the sample in more detail.
Small differences between the reconstructed volumes can be spotted but overall  the NeRF reconstruction has a very good fidelity.

\begin{figure}
  \centering
  \includegraphics[width=\linewidth]{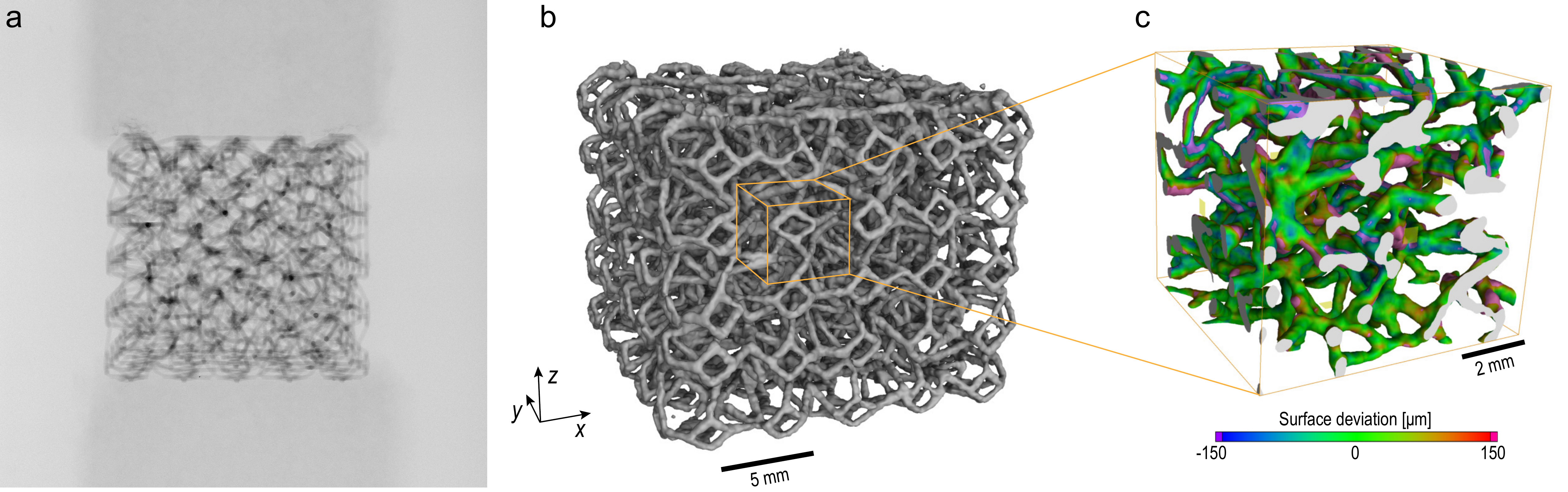}
  \caption{ 
  Reconstruction of randomized Kelvin deformation experiment.
  {(a)} Radiograph at \(t=0\) prior to deformation.
  {(b)} NeRF rendering of the deformed sample at \(t=0.4\).
  {(c)} Detailed view of reconstructed central portion of sample showing surface deviations with respect to withheld X-CT reconstruction.
  } 
  \label{fig:randomized-recon}
\end{figure}

\paragraph{Randomized lattice}
The second experiment with a real 3d printed lattice was designed to test the ability of the framework to reconstruct volumes which do not have nicely ordered periodic structure.
We designed and manufactured a lattice with randomized nodal positions.
The X-ray projection at \(t=0\) is shown in Figure~\ref{fig:randomized-recon}a.
Compared to the radiographs in Figure~\ref{fig:kelvin-recon}a, there is clearly a loss of periodicity of the structure.

For this lattice, the deformation was trained in reverse but the mapping remained linear: \(\Delta / \si{mm} = 5 (1-t)\).
This is because new material enters the volume during compression and our two-component framework needs to be able to map the material seen in projections to the material that is present in the canonical volume. 
Therefore, it is preferred to train the canonical volume on the final step of deformation where the maximum amount of material is present in the field of view.
Figure~\ref{fig:randomized-recon}b shows the NeRF reconstruction at \(t=0.4\). 
A central portion is clipped and showed as an inset in Fig.~\ref{fig:randomized-recon}c.
In this view the loss of ordering and periodicity can be clearly seen.
The surface is colored by surface deviations with respect to withheld high-fidelity X-CT reconstruction.
Note that the resolution of the scan was \SI{75}{\micro m}.
Therefore, most of the reconstructed structure is within 2 voxels of the high-quality X-CT reconstruction.

\section{Discussion and conclusions}
In this work we presented a framework for the spatio-temporal reconstruction of objects from few X-ray projections.
The framework has two main components: a canonical volume, which assigns density to every point in 3d space, and a deformation field which returns a displacement for every point in 3d space as a function of time.
With the two components together, it is possible to query the neural radiance field at continuous temporal and spatial locations.
We train the canonical volume using a hybrid method in which the projection loss is supplemented by volumetric loss based on high-fidelity X-CT data which is typically available at the initial and final timesteps.
The deformation field is trained jointly for all timesteps and with all available data. 
It is posed as a B-spline field whose time-dependent weights are obtained from a neural network.
Such smooth and efficient parametrization makes it feasible to capture the deformation field from just two projections at intermediate timesteps, while not requiring additional loss terms for regularization.
This opens up the possibility for constructing experimental systems with simultaneous acquisition from two source-detector pairs to capture dynamic high-speed deformations in materials.
\footnote{
Since we use a continuous representation of time, the framework can be directly applied in situations when the various dynamic projections are not synchronized but acquired at different times. The angles of the projections can also vary in time and the sample can be rotating as it is being deformed.
}

\paragraph{Applicability}
The work presented here is largely motivated by the goal of reconstructing dynamic material deformations with X-ray tomography.
However, in the first part of the paper, we showed that neural rendering can achieve better reconstruction for static scenes in the sparse-data regime.
Therefore, using neural rendering based on few projections has the potential to speed up static non-destructive X-ray characterization, thus reducing the required time for tomographic data collection from hours to minutes.
Moreover, several key aspects of our work could be used in many other domains, including the more traditional reconstruction of deforming photographic scenes.
For instance, cubic B-spline field with fixed control points and time-varying weights imposes a smooth efficient prior for the deformation field which can be used for the reconstruction of poses.

\paragraph{Limitations}
The key assumption of the two-component framework (canonical model + deformation field) is that deformations preserve topology.
For many real-world deformations, this assumption is not too restrictive and various material behaviour including elasticity, visco-elasticity, and plasticity can be captured with topology-preserving deformations.
However, fracture is an example of a non-topology-preserving deformation.
A different framework, such as full 4d spatio-temporal model, or slicing through higher-dimensional hyperspace, would have to be used.
Some of these methods, however, might require additional loss terms for regularization and it is important to be aware of the inductive biases imposed by many such models (e.g. energy minimization terms often assume isotropic linear elastic material behaviour).

\section*{Acknowledgments}
I.G. is supported by the Ashby PhD Scholarship by the Department of Engineering, University of Cambridge. We thank Vatsa Gandhi for his help with manufacturing the specimens.

\small{
\bibliography{references}
}
\normalsize


\appendix

\section{Appendix}

\subsection{Dataset description} \label{sec:app-datasets}
The domain in which volume is rendered is a cube with side length \num{2} centered at the origin.
Simulated data was generated in an equivalent domain with non-dimensional coordinates \(-1 \leq (x,y,z) \leq 1\).
Experimental data was rescaled appropriately such that the X-CT box coincided with the rendering box with \(-1 \leq (x,y,z) \leq 1\).


\paragraph{Simulated data}
We implemented a custom lightweight ray-tracing algorithm algorithm in \texttt{Go} programming language.
The implementation is linked to our repository and available online.\footnote{\url{https://github.com/igrega348/xray_projection_render/}}
The framework is modular and enables the generation of projections for a variety of objects made of the following primitives: box, cube, parallelepiped, sphere, cylinder.
We implement the option to assemble object collections from objects with positive or negative densities to enable the generation of objects such as shown in Figure~\ref{fig:data-overview}: cube($\rho=1$) + sphere($\rho=-1$) + cylinder($\rho=-1$). 

The framework offers the freedom to choose arbitrary \textit{polar} and \textit{azimuthal} angles $\theta$ and $\varphi$ for the projection. {
For point $\vec{r}=(x,y,z)$, polar angle $\theta$ is defined as the angle between vector $\vec{r}$ and $+z$-axis.
Azimuthal angle $\varphi$ is defined as the angle between the projection of $\vec{r}$ into the $xy$-plane and the $+x$-axis, such that
$\vec{r}=\sqrt{x^2+y^2+z^2}\left( 
\cos{\varphi} \sin{\theta}, \sin{\varphi} \sin{\theta}, \cos{\theta}
\right)$
}
In this work, we use the same data modality for training as is conventional for X-ray tomography: polar angle $\theta=\pi/2$ and azimuthal angles chosen over a range of $2\pi$.
We choose to generate 256 projections uniformly spaced around the circle. 
Different subsets of these projections are later used in training the models.
The test set of images is chosen with randomly varying polar angle in the upper hemisphere \((0 \leq \theta \leq \pi/2) \) according to the rule
$\theta=\arccos{z}$ where $z$ is sampled randomly and uniformly $z \sim U(0,1)$.

The four objects that are used in section \ref{sec:data-efficiency} \textit{(Data efficiency of neural rendering)} are:
\begin{enumerate}
    \item \textit{balls:} 6 spheres distributed pseudo-randomly in space. Three of them have a radius of \num{0.15} while the other three have a radius of \num{0.2}.
    \item \textit{pillars:} 4 cylinders of length \num{1.6} and radius \num{0.1} laid out in a \(2 \times 2\) grid pattern.
    \item \textit{cube:} a cube with a side length of \num{1.5} and an off-center hole which is generated as a cylinder with a spherical cap.
    \item \textit{lattice:} a periodic Kelvin lattice generated as \(4\times 4\times 4\) tessellation of the unit cell. 
    The strut radius is \num{0.025}, and the lattice is between \num{-0.8} and \num{0.8} along all 3 axes.
\end{enumerate}

The lattice used in section \ref{sec:simulated-def-exp} \textit{(Simulated deformation experiment)} is a \(4\times 4\times 4\) Kelvin lattice generated similarly to the lattice above.
We used the aforementioned \texttt{Go} program to remap coordinates during rendering according to the appropriate deformation field.

\paragraph{Experimental data}
X-ray radiographs were acquired using lab-based system Nikon XT H 225ST.
The X-ray settings were adjusted for each sample to provide suitable grey value separation between the material and the background.
Exposure time varied depending on the sample and ranged from \SI{500}{ms} to \SI{4}{s} since the samples were made from different materials and by different processing routes. 
\begin{enumerate}
    \item \textit{balls:} 6 cellulose acetate balls of nominal diameter \SI{2}{mm} inserted into a foam block;
    \item \textit{pillars:} 4 steel rods of nominal diameter \SI{2.5}{mm} and length \SI{23}{mm} inserted into an acrylic box of dimensions \(26\times26\times10\) \si{mm} in a \(2 \times 2\) grid pattern;
    \item \textit{cube:} a plastic cube with a side length of \SI{15}{mm} and an off-center hole with spherical end of radius \SI{3}{mm}, 3d printed using stereolithography (SLA).
    \item \textit{lattices:} various periodic and nonperiodic lattices with an approximate size of \SI{17}{mm} 3d printed using SLA.
\end{enumerate}

In the first part of experiments (Section~\ref{sec:static-reconstructions}), the full dataset consisted of 256 equispaced projections in the $xy$-plane for each object type. 
The in-situ lattice deformation experiments (Section~\ref{sec:real-def-exp}) were done in 50 deformation steps. Full tomography was acquired at deformation steps \(\{0,10,20,30,40,50\}\) and 128 equispaced projections were saved for these steps. For all other deformation steps, 12 equispaced projections were obtained but only 2 were used in this work.

\subsection{B-Spline fields}
\begin{figure}
  \centering
  \includegraphics[width=\linewidth]{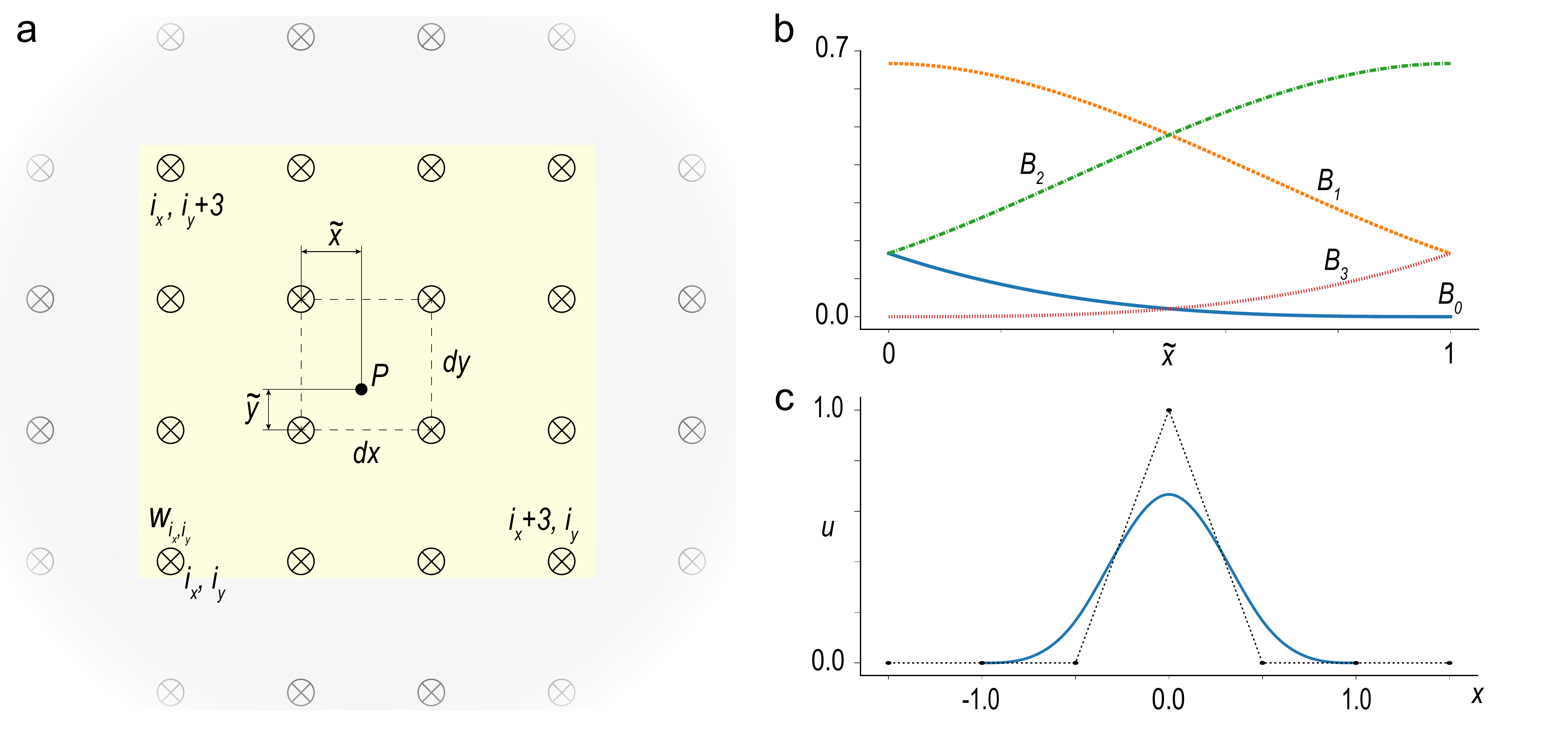}
  \caption{(a) 2d schematic of B-spline field. The grid control points which provide support for query point \(P\) are highlighted in yellow square. Key indices and normalized coordinates are labeled. (b) B-spline base functions. Note \(B_0+B_1+B_2+B_3=1\) for all \(\tx\). (c) Example of B-spline supported function in 1d with weights \([0,0,0,1,0,0,0]\). }
  \label{fig:app-bsplines}
\end{figure}

Here we explain the details of the B-spline method for function interpolation.
Cubic B-splines form a function basis, analogous to Fourier series or other types of decomposition.
The desirable property of B-splines is that the shape functions have \textit{local} influence, much like most finite element (FE) shape functions.
B-spline field is defined by a set of \textit{grid control points} and a set of \textit{weights} corresponding to these control points.
In Figure~\ref{fig:app-bsplines}a, we illustrate the concept with a 2d grid of control points, but in this work we use B-spline fields in 3d and thus we provide equations for the 3d case.
Each control point is indexed by three indices, \(i_x,i_y,i_z\), and has one weight \(w_{i_x,i_y,i_z}\).
The spacing of control points is \(dx, dy, dz\), and the origin of the grid of control points is at \([x_0,y_0,z_0]\).

Let us consider point \(P\) with coordinates \([x,y,z]\) for which we wish to calculate the value of the function \(u\) as given by the B-spline field.
We first need to find the control points which provide \textit{support} to point \(P\).
Indices of the \(4\times4\times 4\) grid of the support points are 
\([i_x,i_{x}+1,i_{x}+2,i_{x}+3]\), 
\([i_y,i_{y}+1,i_{y}+2,i_{y}+3]\) and 
\([i_z,i_{z}+1,i_{z}+2,i_{z}+3]\) where
\[
i_x=\left\lfloor \frac{x - x_0}{dx} - 1 \right\rfloor \qquad 
i_y=\left\lfloor \frac{y - y_0}{dy} - 1 \right\rfloor \qquad 
i_z=\left\lfloor \frac{z - z_0}{dz} - 1 \right\rfloor
\]

The region which is supported by these control points lies in the cube between control points with indices \((i_{x}+1\rightarrow i_{x}+2)\), \((i_{y}+1\rightarrow i_{y}+2)\), \((i_{z}+1\rightarrow i_{z}+2)\), which corresponds to the dashed square in Fig.~\ref{fig:app-bsplines}a.
Normalized coordinates \((\Tilde{x}, \Tilde{y}, \Tilde{z})\) are obtained for within this support region as:
\[
\Tilde{x} = \frac{x - x_0}{dx} - 1 - i_x \qquad \Tilde{y} = \frac{y - y_0}{dy} - 1 -i_y \qquad \Tilde{z} = \frac{z - z_0}{dz} - 1 - i_z
\]
The function value, \(u\), for point \(P\) is then given as a sum of the contributions of all control points weighted by B-spline functions \(B\).
\begin{equation}
    u=\sum_{l=0}^3\sum_{m=0}^3
\sum_{n=0}^3 B_{l}(\Tilde{x})B_{m}(\Tilde{y})B_{n}(\Tilde{z}) w_{i_x+l, i_y+m, i_z+n} \label{eq:Bsplinefield}
\end{equation}

where B-spline functions \(B_k\) are cubic polynomials defined over the range \(0\leq\tx\le1\) which sum to \(1\) and have the form
\begin{align*}
    B_0(\tx) &= \frac{(1-\tx)^3}{6} \\
    B_1(\tx) &= \frac{3 \tx^3-6\tx^2+4}{6} \\
    B_2(\tx) &= \frac{-3\tx^3+3\tx^2+3\tx+1}{6} \\
    B_3(\tx) &= \frac{\tx^3}{6} \\
\end{align*}

We plot these cubic polynomials in Fig.~\ref{fig:app-bsplines}b.
Note that the value of the field at a point which coincides with grid control point 
(\(\tx=0\)) is in general \textbf{not equal} to the value of the weight of that control point.
Such example is shown in one dimension in Fig.~\ref{fig:app-bsplines}c.
The grid has 5 control points with origin \(x_0=-2\) and spacing \(dx=0.5\).
The values of the weights are \([0,0,0,1,0,0,0]\) and the value of the field at location \(x=0\) is \(u(0)=2/3\). Meanwhile, the corresponding control point has weight \(w=1\).

Note that the minimum number of control points for a B-spline field in 3d 
is \(4^3=64\).
To support vector functions, such as displacement, we simply define three sets of weights for the grid, one for each component of the vector field.
Thus, the minimum number of degrees of freedom (weights) for a 3d displacement field is \(3\times64=192\).
The expressive power of the field supported by B-splines increases as the grid of control points is refined and the number of degrees of freedom increases.

\paragraph{Time-varying B-spline field}
In order to represent time-varying displacement, the grid of control points remains fixed and the weights of the field vary with time \(t\).
This is possible in both Eulerian and Lagrangian setting.
\footnote{
In Eulerian setting, displacement value \(u(t,x)\) at time \(t\) and position \(x\) describes the displacement at the corresponding location of space, \(x\), at time \(t\).
In contrast, in Lagrangian setting, the value \(u(t,x)\) describes the displacement \(u\) at time \(t\) of point (particle) which was originally (at time \(t=0\)) at position \(x\).
}
\begin{enumerate}[label=(\alph*)]
    \item We can discretize time into finitely many timesteps, \(t_i\), and define a set of weights at each timestep. The displacement at location \((x,y,z)\) at time \(t\) is then provided by B-spline field supported by weights \(w_{t,i_x+l, i_y+m, i_z+n}\).
    \item Similarly, the field can be defined at discrete timesteps as displacement \textit{increments}, such that \(u(t,x,y,z)=\sum_{t_i<t}\Delta u(t_i,x,y,z)\). 
    The displacement increments are fitted using B-spline field but because Eq.~\eqref{eq:Bsplinefield} is linear in \(w\), 
    the set of weights at time \(t\) is expressible as the sum of all previous weights 
    \(w_t = \sum_{t_i<t}w_{t_i}\).
    \item Both aforementioned approaches could be inefficient for a large number of timesteps. 
    A third approach is to avoid discretizing time into discrete timesteps, but define the set of weights implicitly as a function of time
    \(w_{t,i_x+l, i_y+m, i_z+n} = w_{i_x+l, i_y+m, i_z+n}(t)\). 
    This approach is not only more efficient, but it also builds in implicit time continuity.
\end{enumerate}

We compared the three approaches and obtain the best results with the third approach.

\subsection{Density of B-spline field}
We probe the expressive capability of the B-spline field by an experiment in which two simulated fields with different complexities are applied to \(4\times4\times4\) Kelvin lattice and we attempt reconstruction with B-spline fields with increasing numbers of control points. 

We define \texttt{GaussianDeformation(A,$\sigma$,c)} field parametrized by amplitudes, lengthscales and center, \(A_i, \sigma_i, c_i \), as mapping from undeformed coordinates \(X,Y,Z\) to deformed coordinates \(x,y,z\): 
\begin{align*}
    x &= X + A_0 \exp\left(-\frac{r^2}{2\sigma_0^2}\right) \\
    y &= Y + A_1 \exp\left(-\frac{r^2}{2\sigma_1^2}\right) \\
    z &= Z + A_2 \exp\left(-\frac{r^2}{2\sigma_2^2}\right)
\end{align*}
where \(r^2=(X-c_0)^2+(Y-c_1)^2+(Z-c_2)^2\).

Field 1 consists of a \(2\times2\times2\) grid of \texttt{GaussianDeformation(A,$\sigma=0.2$,c)} where centers are all eight permutations of \(c_i \in \{-0.4,0.4\}\) and amplitudes are \(A_i=[0,0,\pm 0.2]\) with alternating sign. Field 2 follows the same pattern but the grid is \(4\times4\times4\), lengthscale \(\sigma=0.1\), centers are all 64 permutations of \(c_i \in \{-0.6, -0.2, 0.2, 0.6\}\), and amplitudes are \(A_i=[0,0,\pm 0.1]\) with alternating sign.

\begin{figure}
  \centering
  \includegraphics[width=\linewidth]{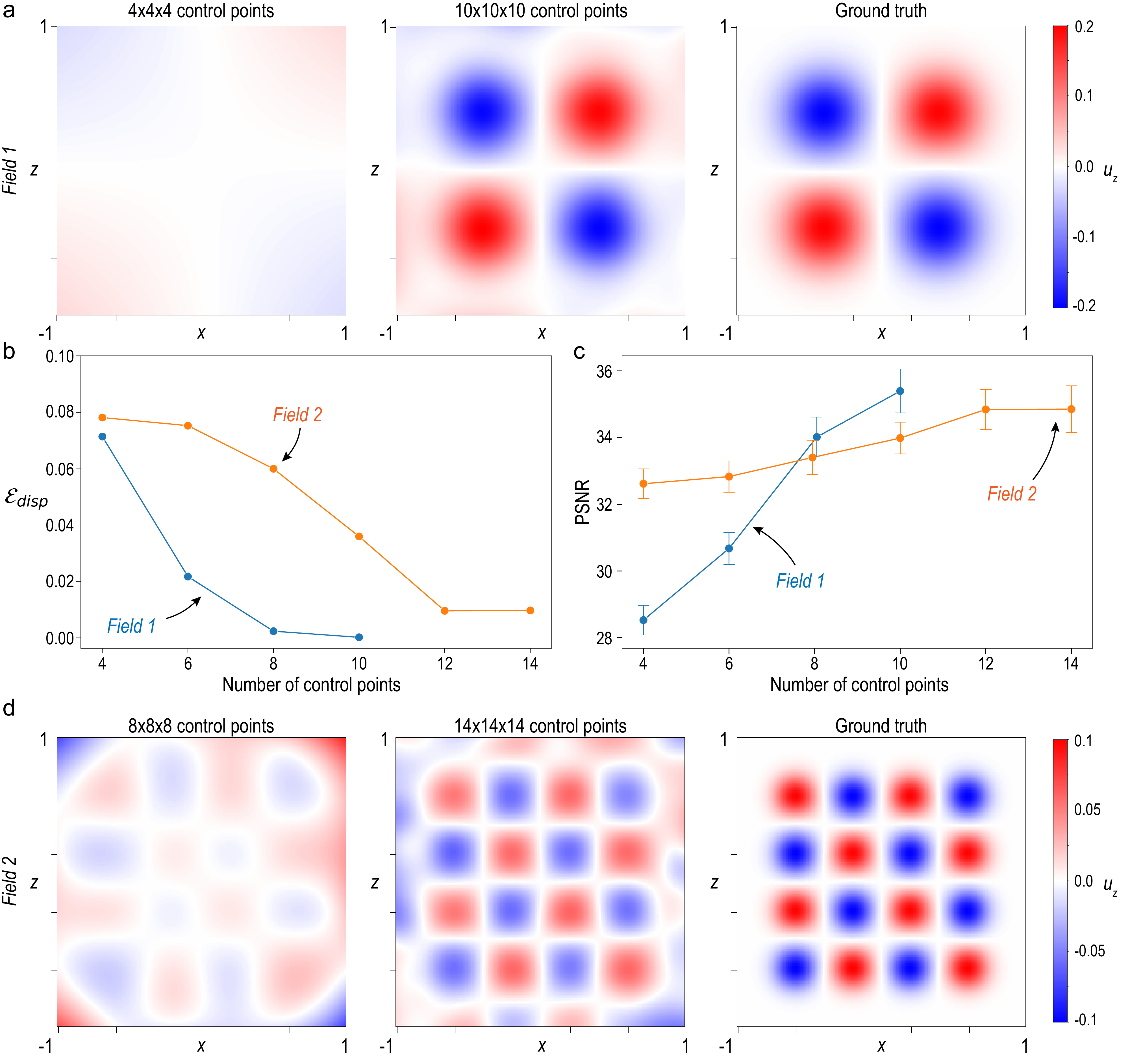}
  \caption{
  \textit{a)} 
  Displacement \(u_z\) at vertical plane \(y=-0.5\) for field 1 (\(2\times2\times2\) grid of Gaussian deformation fields) for two control point densities compared with ground truth.
  \textit{b)}
  Displacement error \(\mathcal{E}_{disp}\) for the two fields as a function of the number of control points.
  \textit{c)} 
  Peak signal to noise ratio (PSNR) for the two fields as a function of the number of control points.
   \textit{d)} 
  Displacement \(u_z\) at vertical plane \(y=-0.6\) for field 2 (\(4\times4\times4\) grid of Gaussian deformation fields) for two control point densities compared with ground truth.
  }
  \label{fig:grid-density}
\end{figure}

Both fields share the same canonical volume. We carry out the reconstruction of the deformation field at various numbers of control points.
In Fig.~\ref{fig:grid-density}a we show the reconstructed displacement \(u_z\) at a vertical plane through the lattice for field 1 (\(2\times2\times2\) grid of Gaussian deformation fields). 
It is clear that the B-spline field with \(4^3\) control points is not sufficient to capture the displacement field, while the B-spline field with \(10^3\) control points achieves an excellent match.
In Fig.~\ref{fig:grid-density}b we plot the displacement error \(\mathcal{E}_{disp}=\frac{1}{|\Omega|} \int_\Omega \| \bm{u}(\bm{r},t) - \bm{\hat{u}}(\bm{r},t) \|^2 \mathrm{d}\Omega\) for the two fields.
We note that as the complexity of the true displacement field increases (field 1 \(\rightarrow\) field 2), a corresponding increase in the complexity of the B-spline field that is used for reconstruction is required.

A natural question to ask is -- how can we choose the required number of control points if we do not have \textit{a priori} knowledge of the deformation field?
In Fig.~\ref{fig:grid-density}c we plot peak signal to noise ratio (PSNR) calculated over validation images which measures the match between the true and rendered projections.
Importantly, the reduction of displacement error \(\mathcal{E}_{disp}\) (not known for an unknown field) is accompanied by an increase in PSNR (observable for an unknown field).
Therefore, the convergence of PSNR can be used to guide the choice of the number of control points for the B-spline field.

\subsection{Trainable flat field} \label{sec:app-flat-field}
We generate two sets of simulated projections for a collection of spheres in 3d: 
\begin{enumerate*}[label=(\arabic*)]
    \item with white background, such that rays passing through empty space (air) do not get attenuated, and 
    \item with grey background, whereby air attenuates the beam and background pixels are not saturated (realistic scenario). 
\end{enumerate*}
We implement trainable flat field in the neural rendering framework. 
Specifically, we can think of decoupling attenuation path into material, \(M\), and air, \(A\):
\[
I=
I_0 \exp\left( - \int_M \mu(s) ds - \int_A \mu(s) ds \right) =
I_0 \exp\left({-\mu_A}\right) \exp\left( - \int_M \mu(s) ds)\right) 
\]
where we use the approximation that the path through air is similar and approximately constant for all rays.
We implement trainable parameter \(f_f\), and derive 
\(\mu_A\) from it using truncation \(\mu_A=\max{(0,f_f)}\) since \(\mu_A\) must not be negative.

\begin{figure}
    \centering
    \includegraphics[width=\linewidth]{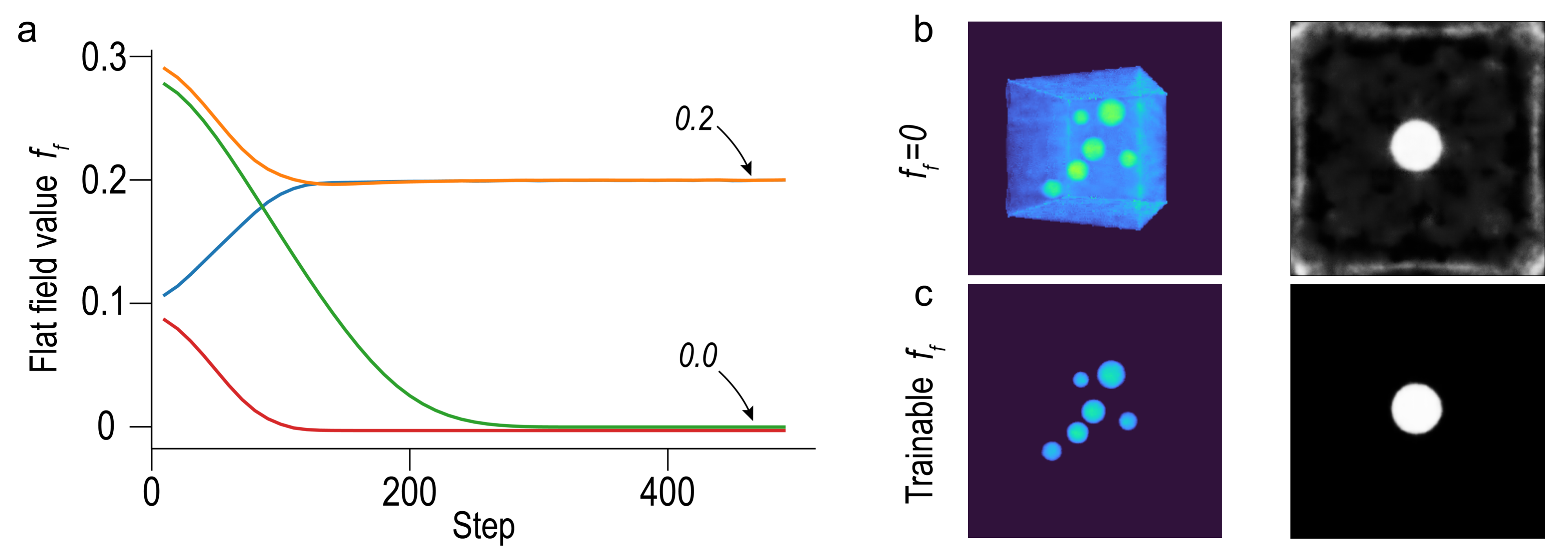}
    \caption{(a) Training flat field value for two sets of projections: white background with flat field value \num{0.0} and grey background with flat field value \num{0.2}.
    For each case, two realizations of initial flat field value are used (\num{0.1}, \num{0.3}). 
    (b,c) 3d accumulation views and cross-sections through trained density field. 
    Non-zero density assigned to nominally empty space can be seen in the accumulation views.
    }
    \label{fig:app-flat-field}
\end{figure}

In Figure~\ref{fig:app-flat-field} we show the convergence of the trained parameter \(f_f\) for both volumes and with \(f_f\) initialized with two values \(f_f\in \{0.1, 0.3 \}\).
The trained flat field \(f_f\) converges to the true value in all cases.
In Fig.~\ref{fig:app-flat-field}b and c we show the benefit of introducing trainable flat field.
When flat field is not trainable but assumed white (b), false density field is obtained. Trainable flat field (c) gives accurate density representation, as it enables the decoupling of material density from background.

\subsection{Information theoretic choice of projection angles}
Here we approach the question "How do we choose the optimum projection angles?" from the perspective of information theory.
Let us define \(\cD\) as a random variable which represents the underlying data in 3d volume.
We observe a fixed number of $N$ 2d projections of this 3d data \(\cX=\cX_1,...,\cX_N\).
Each projection $\cX_i$ has an angle $\theta_i$ associated with it.

\textbf{Postulate 4.1}
\textit{The optimum set of angles $\theta_i$ is such that the entropy of data given all projections \(H(\cD|\cX)\) is minimized:}
\[
H(\cD|\cX) = H(\cD) + H(\cX|\cD) - H(\cX)
\]

The first term, \(H(\cD)\), is the entropy of the data itself and it is determined by the complexity of the 3d object in the volume.
The second term, \(H(\cX|\cD)\),
is the associated with projecting data \cD{} into projections.
If there were no noise, this term would be zero since projections are fully determined by 3d data. 
In presence of noise, the term \(H(\cX|\cD)\) is driven by the level of noise in the projections. 
It can be safely assumed that noise is i.i.d. and therefore projections are conditionally independent of each other given the underlying data \( \cX_i \perp \cX_j | \cD \). 
We can therefore write
\[
H(\cX|\cD) = \sum_i H(\cX_i|\cD) > 0
\]
The term \(H(\cX|\cD)\) is independent of the choice of projection angles \(\theta_i\) and it is detrimental in increasing the entropy of the inferred data \( H(\cD | \cX) \).
The final term \(H(\cX)\) is the joint entropy of projections and it is the one which needs to be maximized in order to reduce the posterior entropy \( H(\cD | \cX) \).

We can decompose the joint entropy \(H(\cX)\) into
\begin{align*}
H(\cX) &= H(\cX_1) + H(\cX_2|\cX_1) + ... + H(\cX_N|\cX_1,...,\cX_{N-1}) \\
&= H(\cX_1) + H(\cX_2) - I(\cX_2;\cX_1) + ... + H(\cX_N) - I(\cX_N;\cX_1,...,\cX_{N-1}) 
\end{align*}
where \(I(\cX;\mathcal{Y})\) is the mutual information between \cX{} and \(\mathcal{Y}\).
Using chain rule, we can successively decompose the higher order mutual information terms as follows:
\begin{align*}
I(\cX_N;\cX_1,...,\cX_{N-1}) &= I(\cX_N;\cX_1) + I(\cX_N;\cX_2,...,\cX_{N-1}|\cX_1) \\
&= I(\cX_N;\cX_1) + I(\cX_N;\cX_2|\cX_1) + I(\cX_N;\cX_3,...,\cX_{N-1}|\cX_1,\cX_2) \\
&= I(\cX_N;\cX_1) + I(\cX_N;\cX_2|\cX_1) + ... + I(\cX_N;\cX_{N-1}|\cX_1,...,\cX_{N-2})
\end{align*}

We now make the following assumption: \textit{Let mutual information be just a function of pairwise interactions between projections, hence} 
\(I(\cX_i;\cX_j|\cX_k)=I(\cX_i;\cX_j)\).
Then, the joint entropy of projections can be written as
\[
H(\cX) = \sum_i H(\cX_i) - \sum_{\substack{1<i<N \\ i<j\leq N}} I(\cX_i; \cX_j)
\]
\textit{To maximize the joint entropy of projections, the sum of all pairwise mutual information terms needs to be minimized.}

\paragraph{Form of pairwise mutual information}
Statistically, for an ensemble of 3d volumes \(\cD\), the mutual information between projections \(\cX_i\) and \(\cX_j\) taken at angles \(\theta_i\) and \(\theta_j\), respectively, should be shift-invariant and just a function of the absolute angular distance between \(\theta_i\) and \(\theta_j\):
\[
I(\cX_i;\cX_j) = I(\theta_i; \theta_j) = I(\theta_i+\Delta;\theta_j+\Delta) \quad \forall \Delta
\]
\[
I(\cX_i;\cX_j)=I(|\theta_j - \theta_i|) =I(\alpha_{ij})
\]
where we define symmetric distance in angular space \(\alpha_{ij}=\alpha_{ji}=|\theta_j - \theta_i|\).

We now reason about the likely form of \(I(\alpha_{ij})\) function.
Because of symmetry and periodicity, function \(I(\alpha_{ij})\) needs to be even and periodic with period \(2\pi\). 
It can therefore be written as Fourier series:
\begin{equation}
I(\alpha_{ij}) = \sum_{k=0}^{\infty} a_k \cos{k \alpha_{ij}} \label{eq:fourier-series}
\end{equation}
Depending on the geometry of X-ray beam:
\begin{enumerate}[label=(\Alph*)]
    \item \textit{Parallel beam:} In case of parallel beams, the projection at \(\theta_2=\theta_1+\pi\) does not provide any more information on top of what is known from the first projection \(\theta_1\). Therefore, \(I(\pi) = I(0)\) and the pairwise mutual information must be periodic with period \(\pi\).
    All odd coefficients in Eq.~\eqref{eq:fourier-series} must be zero.
    \item \textit{Cone beam:} As the perspective angle increases,
    the projection \(\cX_2\) at \(\theta_2=\theta_1+\pi\) does contribute some new information on top of what is known from \(\cX_1\) at \(\theta_1\). The corresponding mutual information should reduce
    \(I(\pi)<I(0)\). Such reduction can be accommodated by non-zero odd coefficients \(a_k (k=1,3,...)\).
\end{enumerate}
In both scenarios, since \(I\) is even, it must be symmetric about \(\alpha_{ij}=\pi\). We expect the function to have a single minimum in the range \(0<\alpha<\pi\).
Since the joint entropy \(H(\cX_i,\cX_j)\) must be non-negative,
the maximum value that \(I(\cX_i;\cX_j)\) can attain is 
\( \max(H(\cX_i),H(\cX_j)) \).

\begin{figure}
    \centering
    \includegraphics[width=\linewidth]{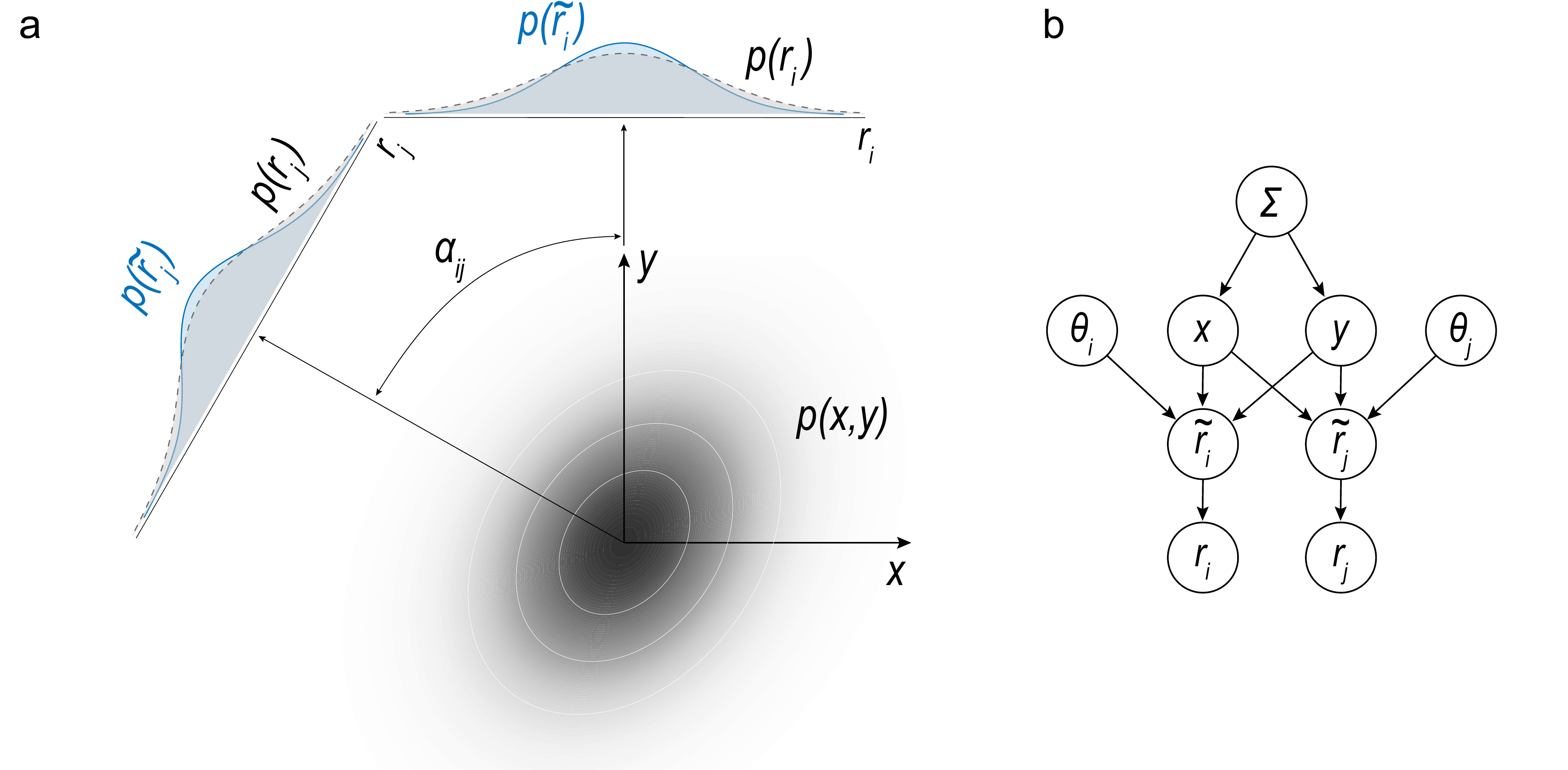}
    \caption{{(a)} Two-dimensional slice with assumed Gaussian probability distribution \(p(x,y)\) with covariance matrix \(\bm{\Sigma}\), noiseless projections (marginals) \(p(\Tilde{r}_i)\) and noisy projections \(p({r}_i)\). 
    Angle \(\alpha_{ij}\) denotes angular separation between the projections.
    {(b)} The assumed conditional probability model.
    }
    \label{fig:app-gaussian}
\end{figure}

\paragraph{Pairwise mutual information for Gaussian density function}
Here we derive the functional form of mutual information between two projections separated by angle \(\alpha_{ij}\) under two simplifying assumptions: 
\begin{enumerate*}[label=(\roman*)]
    \item parallel beam, and
    \item Gaussian distributed data \(\cD\).
\end{enumerate*}.
The geometry of the system can be understood based on the sketch in Fig.~\ref{fig:proj-geom}a.
With parallel beams, all horizontal planes \(x,y\), are decoupled.
We thus only need to consider projections of Gaussian data in 2d plane into 1d slices. 
This is illustrated in Figure~\ref{fig:app-gaussian}a and the tree of the assumed conditional model is in Fig.~\ref{fig:app-gaussian}b.

Let us assume that data \(\cD\) is distributed as Gaussian with zero mean and covariance \(\bm{\Sigma}=\bm{\Sigma}(\sigma_x^2,\sigma_y^2,\rho)\):
\[
\cD \sim p(x,y) \sim \mathcal{N}(\bm{0},\bm{\Sigma})
\]
where coordinates \((x,y)\) are fixed in space.
Consider two projections, \(\cR_i\) and \(\cR_j\) with associated coordinates \((r_i,r_j)\) and angles \((\theta_i,\theta_j)\).
The projections contain noise which is assumed to be additive Gaussian with variance \(\sigma_0^2\).
Noiseless versions of projections are denoted \(\Tilde{\cR}_i\) and \(\Tilde{\cR}_j\), such that 
\(p(r_i|\Tilde{r_i})=\mathcal{N}(r_i,\sigma_0^2)\). 
The coordinate transformation between \((x,y)\) and \((\Tilde{r}_i,\Tilde{r}_j)\) is given by
\[
\Tilde{r}_i=x\cos{\theta_i}+y\sin{\theta_i}
\]

We now proceed to calculate the mutual information between \(\cR_i\) and \(\cR_j\) as a function of the angle of separation \(\alpha_{ij}\).
Mutual information can be calculated as KL-divergence between the joint distribution \((\cR_i,\cR_j)\) and the Kronecker product of \(\cR_i\) and \(\cR_j\) :
\[
I(\cR_i,\cR_j)=D_\mathrm{KL}(P_{(\cR_i,\cR_j)}||P_{\cR_i}\otimes P_{\cR_j})
\]
We need to derive the form of joint probability distribution \(p(r_i,r_j)\) and the form of marginals \(p(r_i),p(r_j)\).
The tree of the assumed conditional model is shown in Fig.~\ref{fig:app-gaussian}b.
Without loss of generality, in the calculations below we shall assume
\(\theta_i=0\), such that \(\alpha_{ij}=\theta_i=\alpha\).
For simplicity, we drop conditioning on \(\alpha_{ij}\) in the notation of probability density functions.
The joint distribution \(p(r_i,r_j)\) can be obtained by marginalization:
\[
p(r_i,r_j)=\int_{x,y,\Tilde{r}_i,\Tilde{r}_j} p(r_i,r_j,\Tilde{r}_i,\Tilde{r}_j,x,y)=
\int_{x,y,\Tilde{r}_i,\Tilde{r}_j} p(r_i|\Tilde{r}_i)p(r_j|\Tilde{r}_j)
p(\Tilde{r}_i|x) p(\Tilde{r}_j|x,y) p(x,y)
\]
The noiseless projection functions lead to Dirac \(\delta\)-functions for \(p(\Tilde{r}_i|x) p(\Tilde{r}_j|x,y)\)
which are straightforward to marginalize over. We thus obtain
\begin{align*}
    p(r_i,r_j)&=\int_{x,y} p(r_i|\Tilde{r}_i=x)p(r_j|\Tilde{r}_j=x\cos{\alpha}+y\sin{\alpha})
p(x,y)    \\
&= \int_{x,y} \mathcal{N}(\Tilde{r}_i;x,\sigma_0^2) \mathcal{N}(\Tilde{r}_j;x\cos{\alpha}+y\sin{\alpha},\sigma_0^2) \mathcal{N}(x,y;\bm{0},\bm{\Sigma}) \\
&= \int_{x,y} \mathcal{N}(r_i,r_j,x,y; \bm{0}, \bm{\Sigma_{rx}}) = \mathcal{N}(r_i,r_j; \bm{0}, \bm{\Sigma_r})
\end{align*} 
where
\[
\bm{\Sigma_{rx}^{-1}}=
\begin{bmatrix}
\frac{1}{\sigma_0^2} & 0 & -\frac{1}{\sigma_0^2} & 0\\
0 & \frac{1}{\sigma_0^2} & -\frac{\cos\alpha}{\sigma_0^2} & -\frac{\sin\alpha}{\sigma_0^2} \\
-\frac{1}{\sigma_0^2} & -\frac{\cos\alpha}{\sigma_0^2} & \frac{1}{\sigma_0^2}+\frac{1}{\sigma_x^2(1-\rho^2)}+\frac{\cos^2\alpha}{\sigma_0^2} & \frac{\rho}{\sigma_x \sigma_y (1-\rho^2)}+\frac{\sin\alpha \cos\alpha}{\sigma_0^2} \\
0 & -\frac{\sin\alpha}{\sigma_0^2} & \frac{\rho}{\sigma_x \sigma_y (1-\rho^2)}+\frac{\sin\alpha \cos\alpha}{\sigma_0^2} & \frac{1}{\sigma_y^2 (1-\rho^2)}+\frac{\sin^2\alpha}{\sigma_0^2}
\end{bmatrix}
\]
\[
\bm{\Sigma_r}=
\begin{bmatrix}
    \sigma_0^2 & \\
    & \sigma_0^2 \\
\end{bmatrix}
 + 
\begin{bmatrix}
    \bm{b_i}^T\bm{\Sigma}\bm{b_i} & \bm{b_i}^T\bm{\Sigma}\bm{b_j} \vspace{1mm}\\
    \bm{b_i}^T\bm{\Sigma}\bm{b_j} & \bm{b_j}^T\bm{\Sigma}\bm{b_j} \\
\end{bmatrix}
\]

The resulting mutual information is
\begin{align*}
I(\cR_i;\cR_j) &= I(\alpha|\bm{\Sigma})= -\frac{1}{2} \ln{\left( 1- \frac{
    \left(\bm{b_i}^T\bm{\Sigma}\bm{b_j} \right)^2}{
    (\sigma_0^2+\bm{b_i}^T\bm{\Sigma}\bm{b_i})(\sigma_0^2+\bm{b_j}^T\bm{\Sigma}\bm{b_j})
    }
    \right)
    }
\end{align*}
where direction vectors \(\bm{b_i}=[ 1, 0 ]^T\), \(\bm{b_j}=[ \cos\alpha, \sin\alpha ]^T\).
If we take the simplest possible case of an axisymmetric \(\Sigma=\mathrm{diag}(\sigma^2)\), the expression reduces to 
\begin{align}
    I(\alpha) = -\frac{1}{2}\ln{\left(1-\frac{\cos^2\alpha_{ij}}{(1+\varepsilon)^2}\right)} \label{eq:mut-inf-simple-app}
\end{align}
where \(\varepsilon=\sigma_0^2/\sigma^2\) is the dimensionless ratio of noise in projection generating process and the variance of the data distribution.

\begin{figure}
  \centering
  \includegraphics[width=\linewidth]{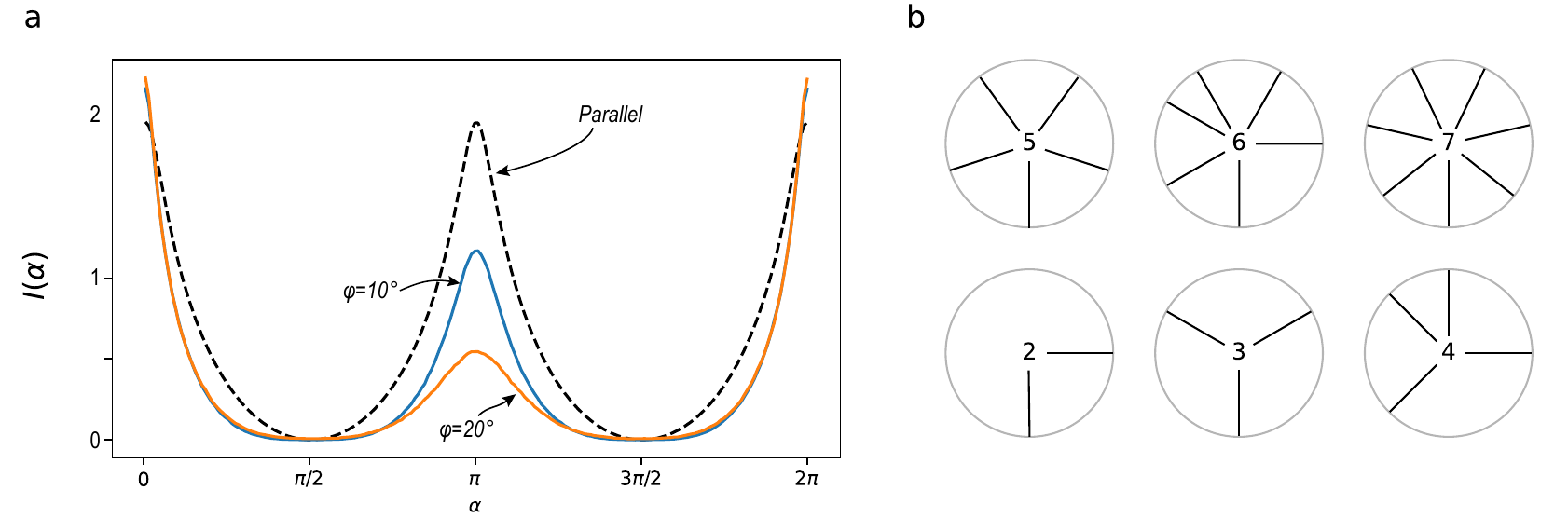}
  \caption{(a) Mutual information as a function of the angle between projections for various cone angles.
  For parallel beams, analytical formula from Eq.~\eqref{eq:mut-inf-simple-app} is plotted while numerical estimates are shown for the other cone angles.
  (b) The optimum angles to take a fixed number of projections.}
  \label{fig:surrogate-projections}
\end{figure}


\paragraph{The effect of cone angle}
Intuitively, we can argue that the effect of cone angle is to decorrelate the projections at 
\(\theta_i\) and \(\theta_j=\theta_i+\pi\).
Mathematically, it is intractable to marginalize over \(p(r_i,r_j,\Tilde{r}_i,\Tilde{r}_j,x,y)\) to derive a closed-form expression for mutual information but we can use the Monte Carlo method.

Let us position the source at distance \(L\) from the origin such that for a projection at angle \(\theta\), the position vector of the source is \(\bm{s}=[L\sin\theta, -L\cos\theta]\).
The imaging plane is positioned opposite the source, also at distance \(L\) from the origin such that its equation is \(\bm{x}.\bm{n}=L\) where the normal vector is \(\bm{n}=[-\sin\theta, \cos\theta]\).
A point with coordinates \(\bm{x}=[x,y]\) is projected on \(\Tilde{r}\) at the imaging plane, where
\begin{equation}
    \Tilde{r} = \left| \left| \bm{x} + (\bm{x}-\bm{s}) \frac{L-\bm{x}.\bm{n}}{(\bm{x}-\bm{s}).\bm{n}} - L \bm{n}  \right| \right|   \label{eq:r-cone} 
\end{equation}

We consider two projections with corresponding source positions \(\bm{s}_1,\bm{s}_2\), and the associated imaging planes with normal vectors \(\bm{n}_1,\bm{n}_2\) while setting \(\theta_1=0\).
Note that the cone angle is implicitly set by the choice of \(L\) relative to the spread of the distribution \(p(x,y)\).
Mutual information \(I(\alpha)\) is calculated as follows:
\begin{enumerate}
    \item For each \(\alpha=\theta_2-\theta_1=\theta_2\), sample \(N=\num{100000}\) points \((x,y)\) from axisymmetric Gaussian with unit variances \(\mathcal{N}(\bm{x};\bm{0};\mathrm{eye}(\bm{1}))\).
    \item Apply coordinate transformation according to Eq.~\eqref{eq:r-cone} to obtain projections \(\Tilde{r}_1, \Tilde{r}_2\).
    \item Add Gaussian noise with \(\sigma_0=0.1\) to \(\Tilde{r}_1, \Tilde{r}_2\) to obtain \({r}_1, {r}_2\).
    \item Calculate correlation coefficient \(\rho\) between \({r}_1\) and \({r}_2\).
    \item Calculate mutual information as \(I(\alpha)=-0.5 \log (1-\rho^2)\)
\end{enumerate}

In Figure~\ref{fig:surrogate-projections}a we plot the resulting mutual information for two cone angles \(\varphi=10^\circ\) and \(\varphi=20^\circ\).\footnote{We use an approximate relationship for cone angle \(\tan(\varphi/2)=\sigma/L=1/L\).}
We note that the projections at \(0^\circ\) and \(\pi\) are indeed decorrelated and \(I(\pi)<I(0)\).
Furthermore, the bigger the cone angle, the more significant this effect.

\paragraph{Minimizing the joint mutual information}
We ultimately wish to calculate optimum projection angles. 
For this reason, it is desirable to have a closed-form expression for pairwise mutual information.
We approximate the curve using Fourier series in the form:
\[I(\alpha) = - \frac{1}{2} \log \left( \sum_{i=0}^{N} a_i \cos\left(i \alpha \right) \right) \]
We tabulate coefficients up to \(\cos6\alpha\) for various cone angles in Table~\ref{tab:fourier} (for a fixed level of noise with \(\varepsilon=\sigma_0/\sigma=0.1\)).

\begin{table}[]
    \centering
    \begin{tabular}{lccccccc}
        \toprule
        \(\varphi [^\circ]\) & \(a_0\) & \(a_1\) & \(a_2\) & \(a_3\) & \(a_4\) & \(a_5\) & \(a_6\) \\
        \midrule 
        10 & 0.66 & -0.02 & -0.46 & -0.01 & -0.13 & 0.0 & -0.01 \\
        15 & 0.67 & -0.05 & -0.43 & -0.03 & -0.12 & -0.01 & -0.01 \\
        20 & 0.69 & -0.10 & -0.39 & -0.05 & -0.10 & -0.01 & -0.02 \\
        25 & 0.73 & -0.16 & -0.29 & -0.07 & -0.08 & -0.02 & -0.02 \\
        \bottomrule \\
    \end{tabular}
    \caption{Tabulated Fourier series coefficients for the surrogate form of pairwise mutual information including the effect of cone angle. See text for full formula.}
    \label{tab:fourier}
\end{table}

Using this surrogate form of mutual information, we numerically optimize projection angles \(\theta_i, i=1,...,N\) for a fixed number of projections, \(N\).
The results of this optimization are sketched in Figure~\ref{fig:surrogate-projections}b. 
The case of odd \(N\) leads to the intuitive result
\(
    \theta_i = 2\pi i / N 
\).
For even \(N\), parallel beams would result in the optimum set of angles \(\theta_i\) dividing half-circle \(\pi\) into equal segments: \(\theta_i = \pi i / N \) .
However, the effect of the cone angle is such that the higher angular separation can be achieved by reflecting some angles into the other half-circle. The effect can be observed in Figure~\ref{fig:surrogate-projections}b.




\end{document}